\documentclass[prd,nofootinbib,twocolumn,a4paper,showkeys,showpacs,preprintnumbers]{revtex4}

\usepackage{graphicx}
\usepackage{amsmath}
\usepackage{amssymb}
\usepackage{amsfonts}
\usepackage{latexsym}
\usepackage{mathrsfs}
\newcommand{\tr}{{\rm tr}}
\newcommand{\Tr}{{\rm Tr}}
\newcommand{\sign}{{\rm sign}}
\renewcommand{\Re}{{\rm Re}}

\newcommand{\dg}{\mathscr{D}^4}
\newcommand{\textint}{\textstyle\int}
\newcommand{\intg}[1]{{\textstyle\int}_{#1}\,}

\newcommand{\CP}{\textit{CP}{}}

\newcommand{\momp}{{p}_2}
\newcommand{\momq}{{p}_3}
\newcommand{\momr}{{p}_4}

\newcommand{\momiabs}{|\vec{p}_i|}
\newcommand{\momabs}{|\vec{p}|}
\newcommand{\momkabs}{|\vec{p}|}
\newcommand{\mompabs}{|\vec{p}_2|}
\newcommand{\momqabs}{|\vec{p}_3|}
\newcommand{\momrabs}{|\vec{p}_4|}

\newcommand{\lorig}{L}
\newcommand{\ltrafo}{{\tilde{L}}}

\newcommand{\carray}[3]{\tilde{C}_{#1}\ifeqthenelse{}{#3}{}{[{#3}]}\ifeqthenelse{}{#2}{}{(#2)}}
\newcommand{\carrayorig}[3]{{C}_{#1}\ifeqthenelse{}{#3}{}{[{#3}]}\ifeqthenelse{}{#2}{}{(#2)}}

\newcommand{\qstat}[1]{1+#1}
\newcommand{\feqtrans}[1]{f^{eq}_a\ifeqthenelse{}{#1}{}{(#1)}}
\newcommand{\fbareqtrans}[1]{\bar{f}^{eq}_a\ifeqthenelse{}{#1}{}{(#1)}}
\newcommand{\hubblerate}{{H}}

\newcommand{\Cmubinitial}{0.5}
\newcommand{\Cmasspsi}{10^{10}}

\newcommand{\Cminrateratio}{10^3}
\newcommand{\Clowerbound}{0.025}
\newcommand{\Cupperbound}{50.0}

\newcommand{\Cdim}{400}
\newcommand{\Creltol}{10^{-8}}

\newcommand{\Ckappaa}{0.01}
\newcommand{\Ckappab}{0.1}
\newcommand{\Ckappac}{0.366}
\newcommand{\Ckappad}{1}
\newcommand{\Ckappae}{10}
\newcommand{\Ckappaf}{100}

\newcommand{\Ckappamaxeta}{0.059}
\newcommand{\Ckappacmaxreleta}{0.34}

\newcommand{\ifeqthenelse}[4]{\edef\tempa{#1}\def\tempb{#2}\ifx\tempa\tempb{#3}\else{#4}\fi}

\newcommand{\momitrans}{{k}_i}
\newcommand{\momktrans}{{k}_1}
\newcommand{\momptrans}{{k}_2}
\newcommand{\momqtrans}{{k}_3}
\newcommand{\momrtrans}{{k}_4}

\newcommand{\enitrans}{{k}^0_i}
\newcommand{\enktrans}{{k}^0_1}
\newcommand{\enptrans}{{k}^0_2}
\newcommand{\enqtrans}{{k}^0_3}
\newcommand{\enrtrans}{{k}^0_4}

\newcommand{\massitrans}{{m}_i}

\newcommand{\massptrans}{{m}_2}

\newcommand{\dif}{d}
\newcommand{\ftrans}[2]{{\tilde{f}}_{#1}}

\newcommand{\heaviside}[1]{\theta\big(#1\big)}
\newcommand{\abs}[1]{\left|#1\right|}

\newcommand{\rhobarc}{\rho_C}
\newcommand{\rhobarf}{\rho_f}
\newcommand{\hab}{{h}}
\newcommand{\fpsitrans}{f_{\psi_1}}

\renewcommand{\vec}[1]{{\bf #1}}

\usepackage{color}
\definecolor{blue}{rgb}{0,0,0.5}
\definecolor{lightblue}{rgb}{0,0,1}
\definecolor{red1}{rgb}{1,0,0}
\definecolor{red}{rgb}{0.5,0,0}
\definecolor{lightred}{rgb}{1,0.5,0}
\definecolor{green}{rgb}{0,0.5,0}
\definecolor{darkgreen}{rgb}{0.0,0.3,0.0}
\definecolor{grey}{rgb}{0.5,0.5,0.5}

\begin{document}
\preprint{TUM-HEP-735/09}

\title{Systematic approach to leptogenesis in nonequilibrium QFT:
\\ vertex contribution to the \textit{CP}-violating parameter}

\author{M. Garny$^{a,b}$}
\email[\,]{mathias.garny@ph.tum.de}

\author{A. Hohenegger$^{a}$}
\email[\,]{andreas.hohenegger@mpi-hd.mpg.de}

\author{A. Kartavtsev$^{a}$}
\email[\,]{alexander.kartavtsev@mpi-hd.mpg.de}  

\author{M. Lindner$^{a}$}
\email[\,]{manfred.lindner@mpi-hd.mpg.de}

\affiliation{%
$^a$Max-Planck-Institut f\"ur Kernphysik, Saupfercheckweg 1, 69117 Heidelberg, Germany\\
$^b$Technische Universit\"at M\"unchen, James-Franck-Stra\ss e, 85748 Garching, Germany}

\begin{abstract}
The generation of a baryon asymmetry via leptogenesis is usually studied
by means of classical kinetic equations whose applicability to processes
in the hot and expanding early universe is questionable. The approximations 
implied by the state-of-the-art  description can be tested in a first-principle
approach based on nonequilibrium field theory techniques. Here, we apply
the Schwinger--Keldysh/Kadanoff--Baym formalism to a simple toy model of leptogenesis. 
We find that, within the toy model, medium effects increase the vertex 
contribution to the \textit{CP}-violating parameter. At high temperatures
it is a few times larger than in vacuum and asymptotically reaches the vacuum
value as the temperature decreases. Contrary to the results obtained earlier
in the framework of thermal field theory, the corrections are only linear 
in the particle number densities. An important feature of the  Kadanoff--Baym formalism 
is that it is free of the  double-counting problem, i.e.~no need for real 
intermediate state subtraction arises.  In particular, this means that the 
structure of the equations automatically ensures that the asymmetry vanishes
in equilibrium. These results give a first glimpse into a number of new and
interesting effects that can be studied in the framework of nonequilibrium field theory.
\end{abstract}

\maketitle

\section{\label{introduction}Introduction}
The almost complete absence of  antimatter on Earth, in the solar system and 
in hadronic cosmic rays suggests that the universe is baryonically asymmetric. 
This conclusion is confirmed by experimental data on the abundances of the light 
elements \cite{Kolb:1990vq} and precise measurements of the cosmic
microwave background spectrum \cite{Hinshaw:2008kr,Komatsu:2008hk}. 

The baryon asymmetry of the universe can be generated dynamically provided
the three Sakharov conditions~\cite{Sakharov:1967dj} are fulfilled in the early 
universe: violation of baryon (or baryon minus lepton) number; violation of 
\textit{C} and \textit{CP}; and deviation from thermal equilibrium.
In the standard model supplemented by heavy right-handed Majorana
neutrinos, these conditions are naturally satisfied for leptons.
The Majorana mass term violates lepton number by two units. Complex 
Yukawa couplings of the right-handed neutrinos to leptons and the Higgs 
doublet induce \textit{CP} violation. The  rapid expansion of the universe 
causes a deviation from thermal equilibrium. Finally, the generated 
lepton asymmetry is converted to the observed baryon asymmetry 
by sphalerons \cite{'tHooft:1976up,Kuzmin:1985mm}. In other words the 
generation of the baryon asymmetry -- baryogenesis -- proceeds 
via the generation of a lepton asymmetry -- leptogenesis \cite{Fukugita:1986hr}.

Many aspects of leptogenesis have been extensively investigated. In particular,
it has been shown that the \textit{CP}-violating parameter and the efficiency of 
leptogenesis are affected by the flavor structure of the neutrino Yukawa couplings 
\cite{Nardi:2006fx,Abada:2006fw,Abada:2006ea, Barbieri:1999ma,Blanchet:2006be,
Blanchet:2007hv,Anisimov:2007mw,Antusch:2006gy,Antusch:2006cw}.
In \cite{Pilaftsis:2003gt,Pilaftsis:2005rv} it was demonstrated that the
\textit{CP}-violating parameter is resonantly enhanced if two of the heavy 
neutrinos have mass differences comparable to their decay widths. Medium
effects have also been addressed. In the hot and dense plasma the deviation
of the \textit{CP}-violating parameter and the thermal masses from their vacuum
values plays an important role \cite{Covi:1997dr,Giudice:2003jh}.
In state-of-the-art calculations Boltzmann equations are used to compute the 
asymmetry. Their applicability in the hot and expanding early universe can be checked 
using a first-principle approach like the Schwinger--Keldysh/Kadanoff--Baym formalism.
Some aspects of leptogenesis have been investigated within this framework at different 
levels of approximation in \textit{Minkowski} space  \cite{Buchmuller:2000nd,
DeSimone:2007rw,DeSimone:2007pa,DeSimone:2008ez,Anisimov:2008dz}. 
These studies were motivated by the expectation that in the expanding universe 
filled with hot and dense plasma quantum effects, which are neglected in the 
canonical treatment, might play a crucial role.

In this paper we investigate leptogenesis, and in particular the 
vertex contribution to the \textit{CP}-violating parameter within the framework 
of nonequilibrium quantum field theory. Using the Kadanoff--Baym formalism 
as the starting point of our analysis, we derive \emph{quantum-corrected Boltzmann equations}. 
We explicitly take medium corrections to the \textit{CP}-violating parameter as well as the
expanding background into account. The Kadanoff--Baym approach to the analysis of 
nonequilibrium systems is technically considerably more involved than the canonical 
Boltzmann ansatz. For this reason, before applying it to realistic models of leptogenesis, 
here we study a simple toy model containing one complex and two real scalar fields. 
It is defined by the Lagrangian
\begin{align}
	\label{lagrangian}
	{\cal L}&=\frac12 \partial^\mu\psi_i\partial_\mu\psi_i
	-\frac12 M^2_i \psi_i\psi_i +\partial^\mu \bar{b}\partial_\mu b-
	m^2\bar{b}b\nonumber\\
	&-\frac{\lambda}{2!2!}(\bar{b}b)^2-
	\frac{g_i}{2!}\psi_i bb-\frac{g^*_i}{2!}\psi_i \bar{b}\bar{b}
	+{\cal L}_{rest}\,, i=1,2\,,
\end{align}
where $\bar{b}$ denotes the complex conjugate of $b$.
Despite its simplicity, the model incorporates all features relevant 
for leptogenesis. The real scalar fields imitate the (two lightest) 
heavy right-handed neutrinos, whereas the complex scalar field models 
the baryons. The $U(1)$ symmetry, which we use to define ``baryon'' number, 
is explicitly broken by the presence of the last two terms, just as the 
$B-L$ symmetry is explicitly broken by Majorana mass terms in 
phenomenological models. Thus the first Sakharov condition is fulfilled.
The couplings $g_i$ model the complex Yukawa couplings of the 
right-handed neutrinos to leptons and the Higgs. By rephasing the 
complex scalar field at least one of the couplings $g_i$ can be made real. 
If ${\rm arg}(g_1)\neq {\rm arg}(g_2)$ the other one remains complex and 
there is  \textit{CP} violation, as is required by the second Sakharov condition.
In vacuum the vertex contribution to the \textit{CP}-violating parameter is given by
\begin{align}
	\label{epsilonvac}
	\epsilon_i^{\it vac}=-&\frac1{8\pi}\frac{|g_j|^2}{M_i^2}{\rm Im}\left(\frac{g_ig^*_j}{g^*_ig_j}\right)
	\ln\biggl(1+\frac{M_i^2}{M_j^2}\biggr)\,,
\end{align}
see Appendix\,\ref{cpclassic}.
Just as in realistic models, the required deviation from thermal equilibrium 
is caused by the rapid expansion of the universe. Thus the third Sakharov condition 
is fulfilled as well. Finally, the quartic self-interaction term in \eqref{lagrangian} plays the  
role of the Yukawa and gauge interactions in established models -- it brings the 
``baryons'' to equilibrium. The renormalizability of the theory requires the presence 
of some additional terms, which are accounted for by ${\cal L}_{rest}$. 
By appropriately choosing the corresponding coupling constants we can always 
make the contributions of these terms  negligibly small. Since the physically interesting 
range for the generation of the asymmetry is $0.1  M_i \lesssim T \lesssim 10 M_i$, 
where $M_i$ is the mass of the lightest heavy particle, the running effects cannot 
make these couplings large during the relevant period. 

Apart from the vertex contribution~\cite{Fukugita:1986hr} to the \textit{CP}-violating 
parameter discussed above, there is also a self-energy contribution~\cite{Flanz:1994yx,Covi:1996wh,PhysRevD.56.5431}.
In the Kadanoff--Baym formalism the analysis of the former is rather independent from 
the analysis of the latter. For this reason, in this paper, we consider only the vertex contribution,
whereas the self-energy contribution will be addressed in \cite{Garny:2009qn}.
To make the discussion more transparent we give the technical details 
in the appendices, whereas in the main body of the paper we discuss 
qualitative features of the employed approach and present the results.
\begin{itemize}
	\item[(i)] As we argue in Sec.\,\ref{KBandB}, the formalism is free of the 
	double-counting problem typical for the canonical Boltzmann approach. 
	In other words the structure of the equations automatically ensures 
	that the asymmetry vanishes in thermal equilibrium and no need for the 
	real intermediate state subtraction arises. 
	\item[(ii)] Our result for the vertex contribution to the \textit{CP}-violating 
	parameter, presented in Sec.\,\ref{CPviol}, differs from that 
	obtained in the framework of equilibrium thermal field theory 
        by replacing the zero-temperature propagators with finite temperature 
	propagators in the matrix elements of the Boltzmann equation
	\cite{Giudice:2003jh,Covi:1997dr} -- the 
	medium corrections are only \textit{linear} in the particle number 
	densities. For scalars the medium effects always increase the 
	\textit{CP}-violating parameter, which in turn leads to an enhancement 
	of the generated asymmetry.
	\item[(iii)] By comparing the \textit{CP}-violating parameters obtained by
	using the Maxwell--Boltzmann (MB) and Bose--Einstein (BE) statistics, we find
	that quantum statistical effects play a considerable role. 
	As we argue in Sec.\,\ref{CPviol}, the medium effects increase the
	\textit{CP}-violating parameter by a factor of at most two in the
	Maxwell--Boltzmann approximation. At high temperatures,
	the increase is up to an order of magnitude larger when Bose enhancement 
	is taken into account. 
\end{itemize}
In Sec.\,\ref{Numerics} we present numerical solutions of the quantum-corrected 
Boltzmann equations, and discuss the quantitative impact of medium effects on the final 
asymmetry within the toy model. Finally, in Sec.\,\ref{Summary}, we summarize our results 
and present our conclusions.

\section{\label{KBandB}Nonequilibrium dynamics}
To calculate the asymmetry generated at the epoch of leptogenesis 
one usually employs generalized Boltzmann equations for the one-particle 
distribution functions of the different particle species \cite{Kolb:1990vq}:
\begin{align}
	\label{blzmnclassic}
	&p^\alpha {\cal D}_\alpha f_\psi(X,p)={\textstyle\frac12}\textint d\Pi_a^3 d\Pi_b^3
	\ldots d\Pi_i^3 d\Pi_j^3 \ldots\nonumber\\
	& \times (2\pi)^4 
	\delta(p+p_a+p_b\ldots-p_i-p_j)\nonumber\\
	&\times [|M|^2_{i+j+\ldots\rightarrow \psi+a+b\ldots}
	f_i f_j\ldots (1\pm f_a)(1\pm f_b)(1\pm f_\psi)\nonumber\\
	&-|M|^2_{\psi+a+b\ldots\rightarrow i+j+\ldots}
	f_a f_b f_\psi \ldots (1\pm f_i)(1\pm f_j)\,.\,.].
\end{align}
The fourth line in Eq.\,\eqref{blzmnclassic} describes the decrease 
in number of species $\psi$ due to the scattering (or decay) process 
$\psi+a+b\ldots\rightarrow i+j+\ldots$ and is usually referred to 
as the loss term. The third line describes the increase in the number 
of $\psi$ due the process $i+j+\ldots\rightarrow \psi+a+b\ldots$ and 
is referred to as the gain term. The Dirac $\delta$ function in the second
line enforces energy-momentum conservation in each individual process,
whereas the invariant phase-space elements $d\Pi^3$ ensure that the
resulting expression is a Lorentz scalar. The probabilities of the 
decay and scattering processes are usually calculated in
\textit{vacuum}, which is inconsistent with the nonzero particle 
number densities. Moreover the canonical approach is plagued with 
the \textit{double-counting} problem. For instance, in the canonical approach, 
the scattering process $\ell h\rightarrow \tilde{\psi}_i \rightarrow \bar{\ell}\bar{h}$ 
is equivalent to the inverse decay ($\ell h, \bar{\ell}\bar{h}\rightarrow \psi_i$) of 
the heavy Majorana neutrino followed by the decay 
($\psi_i\rightarrow \ell h, \bar{\ell}\bar{h}$) if the intermediate heavy neutrino 
$\tilde{\psi}_i$ is on-shell. That is, the same contribution is counted twice. 
As a consequence, a nonzero  asymmetry is generated even in thermal equilibrium. 
The problem is accounted for by the \textit{real intermediate state} 
subtraction procedure. Since the scattering amplitude is calculated in 
\textit{vacuum}, one cannot assign a distribution function to the heavy neutrino $\tilde{\psi}_i$. 
For this reason the resulting collision terms are difficult to interpret. 
Only by assuming Maxwell--Boltzmann statistics and by integrating the 
resulting Boltzmann equations can one derive rate equations
that manifestly lead to zero asymmetry in equilibrium.\footnote{Strictly 
speaking, the system (\ref{blzmnclassic}) is incorrect in general if it involves 
unstable particles, because the matrix elements, computed in perturbation theory, 
do not necessarily meet the requirements for transition amplitudes in systems of 
Boltzmann equations. In cases such as the present (leptogenesis) subtle 
modifications (RIS subtraction) are necessary to obtain consistent results.}

Boltzmann equations rely on the concept of on-shell particles with 
constant mass. The spectral function ${G}_{\rho}(X,p)$ of 
a particle whose mass does not change as it propagates along a geodesic  
is orthogonal to its  four-momentum \cite{Hohenegger:2008zk}:
\begin{align}
	\label{SpctrFuncEq}
	p^\alpha{\cal D_\alpha} {G}_{\rho}(X,p) =0\,.
\end{align}
For a pointlike on-shell particle the spectral function is zero 
off-shell and diverges on-shell:
\begin{align}
	{G}_{\rho}(X,p) = 2\pi \,\sign(p_0)\delta(p^\alpha p_\alpha-M^2)\,.
\end{align}
Instead of using one-particle distribution functions, as the quantities describing the statistical properties of the system, we can use the statistical propagator ${G}_{F}(X,p)$:
\begin{align}
	{G}_{F}(X,p)&\equiv\left[f_\psi(X,p)+{\textstyle\frac12}\right]
	{G}_{\rho}(X,p)\,.
\end{align}
Written in terms of ${G}_{\rho}(X,p)$ and ${G}_{F}(X,p)$, 
a Boltzmann equation takes the form
\begin{align}
	\label{blzmnnewform}
	p^\alpha {\cal D}_\alpha &{G}_{F}(X,p)\nonumber\\
	=&{\textstyle\frac12} [{\Pi}_{<}(X,p){G}_{>}(X,p)-
	{G}_{<}(X,p){\Pi}_{>}(X,p)]\,,
\end{align}
where ${G}_\gtrless\equiv{G}_F\pm \frac12{G}_\rho$. 
Comparing Eqs.\,\eqref{blzmnnewform} and \eqref{blzmnclassic} 
we see that the quantities ${\Pi}_\gtrless$ correspond 
to the loss and gain terms. 

The transformations made so far simply seem to amount to a change 
of notation. However, the situation changes when one realizes that 
Eqs. \eqref{SpctrFuncEq} and \eqref{blzmnnewform} coincide with
the equations that can be derived in a certain approximation from 
the system of Kadanoff--Baym equations and that the quantities ${\Pi}_\gtrless$ can be identified with the self-energies. Despite the close similarity
there is an important difference: the self-energies calculated in 
the framework of the Kadanoff--Baym formalism differ from 
the gain and loss terms obtained in the canonical approach.
As we will show in the following, the self-energies consistently  
take medium effects into account and the resulting equations are free of the double-counting problem.

The Kadanoff--Baym formalism may be viewed as \emph{top-down}
approach: starting from the complete evolution equations for the two-point
functions, it is possible to derive kinetic equations in a systematic way
by applying a number of well-known approximations 
\cite{Danielewicz:1982kk,Ivanov:1999tj,Knoll:2001jx,Blaizot:2001nr}.
We will refer to these equations as \emph{quantum-corrected Boltzmann equations}.
Within the Kadanoff--Baym formalism, both the overall structure
of the equations as well as the scattering and decay rates  are derived
self-consistently from a common starting point based on the \emph{in-in} or
Schwinger--Keldysh description of nonequilibrium quantum fields.
In contrast to that, within the canonical \emph{bottom-up} approach, the 
scattering and decay rates are extracted from the S-matrix (i.e.~based on 
the \emph{in-out} formalism), and are then inserted into Boltzmann equations.
We expect that both approaches are equivalent in the zero-temperature/zero-density
and small coupling limit, where the mean-free path is large compared to the
microscopic interaction length scales.
In this regime, it may be considered that the top-down approach discussed
here adds support to the usual bottom-up formalism. Additionally, as 
mentioned above, the top-down approach is free of the double-counting 
problem and allows one to explore the medium effects within nonequilibrium 
quantum field theory. This is especially relevant for the \CP-violating terms, 
since these contain loop diagrams.

\subsection{Kadanoff--Baym equations}
The system of Kadanoff--Baym equations for the spectral function 
and the statistical propagator is usually formulated in
coordinate space. For the complex scalar field $b$, which  corresponds to
baryons in our model they read (see Appendix\,\ref{Complscal})
\begin{subequations}
	\label{Dequations}
	\begin{align}
		\label{DFequation}
		[\square_x+m^2(x)]D_F(x,y)&=
		{\textint\limits^{y^0}_0} \dg z\,\Sigma_F(x,z)D_\rho(z,y)\nonumber\\
		&-{\textstyle\int\limits^{x^0}_0}\dg z\,\Sigma_\rho(x,z)D_F(z,y)\,,\\
		\label{Drhoequation}
		[\square_x+m^2(x)]D_\rho(x,y)&=
		{\textint\limits_{x^0}^{y^0}}\dg z\,\Sigma_\rho(x,z)D_\rho(z,y)\,.
	\end{align}
\end{subequations}
As is clear from the terminology, the spectral function $D_\rho$
contains information about spectral properties of the system, 
whereas the statistical propagator $D_F$ contains information about
its state. The spectral and statistical self-energies, $\Sigma_\rho$
and $\Sigma_F$, as well as the effective mass $m^2(x)=m^2+\Sigma_{\it loc}(x)$, 
which contains the local self-energy, carry information about the interactions
in the system. They describe scattering and mean-field effects, respectively.
The invariant volume element,
\begin{align*}
	\dg z\equiv \sqrt{-g} d^4 z\,,\quad g\equiv \det g_{\mu\nu}\,,
\end{align*}
ensures that the Kadanoff--Baym equations \eqref{Dequations} can be applied 
to the analysis of out-of-equilibrium dynamics not only in 
Minkowski, but also in a general curved space-time. Particularly 
interesting for our purpose is the case of the expanding early universe 
\cite{Tranberg:2008ae}.

Let us list some of the qualitatively important features. First of all, Eqs.\,\eqref{Dequations} 
are written in terms of \textit{resummed} propagators, i.e.~they take into 
account the full series of \emph{daisy} and \emph{ladder} diagrams (see e.g. 
\cite{Blaizot:2003an}). Second, the characteristic \emph{memory} integrals on 
the right-hand side integrate over the full time history of the evolution. 
In other words the Kadanoff--Baym equations are non-Markovian, 
i.e.~are not local in time.
It is very important that the Kadanoff--Baym equations do not rely on 
the concept of quasiparticles and their collisions in the plasma. 
In other words, they are free of any possible uncertainties associated 
with definition of quasiparticle excitations in the hot plasma of
the rapidly expanding universe. This property makes the Kadanoff--Baym
equations a prime candidate for the analysis of leptogenesis. If 
the quasiparticle picture is applicable, they account for the 
time-dependence of the quasiparticle parameters. In particular, 
an effective time-dependent mass and width induced by the interactions 
of the system can be extracted from the Wigner-transform of the spectral 
function~\cite{Aarts:2001qa}. Finally, for weakly coupled 
systems close to thermal equilibrium the Kadanoff--Baym equations 
can be reduced to the Boltzmann 
equation~\cite{Danielewicz:1982kk,Ivanov:1999tj,Knoll:2001jx,
Blaizot:2001nr}, which we have briefly discussed above.

The Kadanoff--Baym equations for a system of $n$ real scalar fields read 
\cite{Hohenegger:2008zk,Garny:2009qn}
	\begin{subequations}
	\label{Gequations}
	\begin{align}
		\label{GFequation}
		[\square_x+M_i^2]G^{ij}_{F}(x,y)&=
		{\textint\limits^{y^0}_0}\dg z\,\Pi^{ik}_{F}(x,z)G^{kj}_{\rho}(z,y)\nonumber\\
		&-{\textstyle\int\limits^{x^0}_0}\dg z\,\Pi^{ik}_{\rho}(x,z)G^{kj}_{F}(z,y),\\
		\label{Grhoequation}
		[\square_x+M_i^2]G^{ij}_{\rho}(x,y)&=
		{\textint\limits_{x^0}^{y^0}}\dg z\,\Pi^{ik}_{\rho}(x,z)G^{kj}_{\rho}(z,y),
		\hskip -1mm
	\end{align}
\end{subequations}
where the propagators and self-energies are now $n$--by--$n$ matrices. 
The off-diagonal components of the propagators and self-energies 
describe the mixing of the fields.

\begin{figure}[!ht]
	\begin{center}
		\includegraphics[width=0.48\textwidth]{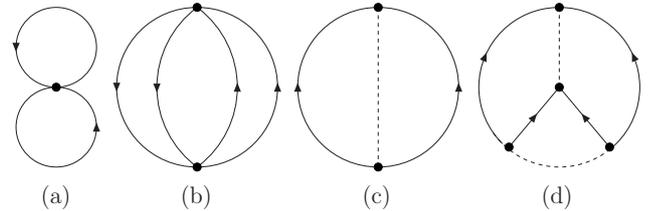}
	\end{center}
\caption{\label{diagrams}Relevant two-- and three-loop contributions 
to the 2PI effective action.}
\end{figure}

Equations \eqref{Dequations} and \eqref{Gequations} are valid for a system of 
one complex and $n$ real scalar fields with arbitrary interactions. Here, we consider $n=2$.
The information about the particular form of the interactions is encoded 
in the corresponding self-energies $\Sigma(x,y)$ and $\Pi^{ij}(x,y)$. 
The latter ones can be derived by functional differentiation of the 
two--particle--irreducible (2PI) effective action, $\Gamma_2[G,D]$, with 
respect to the two-point correlation functions $D(y,x)$ and $G^{ji}(y,x)$,
see \cite{Hohenegger:2008zk} and Appendix\,\ref{Complscal} for more details. 
The effective action is given by the sum of all 2PI diagrams with 
vertices as given by the interaction Lagrangian and internal lines 
representing the complete connected two-point functions \cite{Berges:2004yj}. 
Of course the number of 2PI diagrams contributing to the effective
action is infinite in any theory and the infinite sum must be truncated
in a suitable way. To obtain a qualitative similarity
between the toy model and the established models, we must take
into account processes which generate and washout the asymmetry. 
The minimal set of relevant 2PI contributions is presented in Fig.\,\ref{diagrams}.
Clearly, after cutting any two lines of the diagrams $(a)-(d)$ they 
still remain connected. To understand which physical processes are 
described by the above diagrams in the Boltzmann approximation one
has to discriminate between local and nonlocal contributions.
\begin{figure}[!ht]
	\begin{center}
		\includegraphics[width=0.42\textwidth]{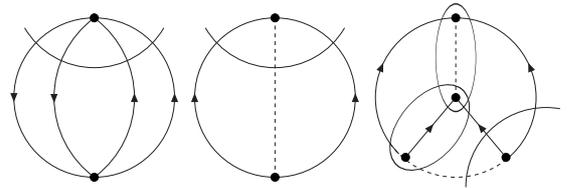}
	\end{center}
\caption{\label{processes}Processes described by the 2PI diagrams in Fig.\,\ref{diagrams}.}
\end{figure} 
The local diagram $(a)$ generates the mean-field correction 
$\Sigma^{\rm loc}(x)$ to the effective mass $m^2(x)$ of the field:
\begin{align}
	\Sigma^{\rm loc}(x)=\lambda D(x,x) \; .
\end{align}
The nonlocal diagrams $(b)$--$(d)$ describe scattering and decay 
processes, which can be identified by cutting the diagrams into two 
pieces by drawing a connected line in all possible ways as indicated in Fig.\,\ref{processes}
(see also \cite{Carrington:2004tm}).
Cutting diagram $(b)$, we obtain  squares of tree-level
amplitudes of $bb\rightarrow bb$ and $b\bar{b}\rightarrow b\bar{b}$
scattering processes. Analogously, cutting  diagram $(c)$, we get 
squares of the tree-level amplitudes of  $\psi_i\rightarrow bb$
and $\psi_i\rightarrow \bar{b}\bar{b}$ decay processes. In the canonical approach
the decays of the heavy real scalars are \textit{CP}-conserving at tree-level.
To leading order the vertex \textit{CP}-violating parameter $\epsilon$ 
is generated by interference of the tree-level and one-loop vertex
decay amplitudes. 
\begin{figure}[!ht]
	\begin{center}
		\includegraphics[width=0.45\textwidth]{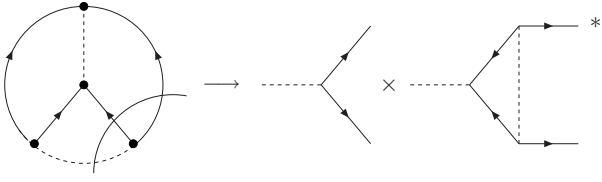}
	\end{center}
\caption{\label{interference}Interference of the tree-level and one-loop 
vertex decay amplitudes.}
\end{figure} 
In the Kadanoff--Baym formalism the leading vertex 
\textit{CP}-violating contribution is described by diagram $(d)$. 
Cutting it in the way presented in Fig.\,\ref{interference}, we obtain the 
product of the tree-level and one-loop vertex amplitudes.\footnote{In addition, 
there are two other ways to cut this diagram, 
which are denoted by the ellipses in Fig.\,\ref{processes}. 
They describe scattering processes $bb\rightarrow \bar{b}\bar{b}$, 
$b\bar{b}\rightarrow \psi_i\psi_j$ and $\psi_i b\rightarrow \psi_j b$. 
Note that the three-loop 2PI diagram only describes the interference of the 
$s$, $t$ and $u$--channel scattering amplitudes: $M_{st}$, $M_{su}$,
$M_{tu}$. The missing topologies, which generate the $M_{ss}$, $M_{tt}$,
and $M_{uu}$ terms, appear only upon use of the \textit{extended 
quasiparticle approximation}. This analysis is beyond the scope of
this paper and will be presented in \cite{Garny:2009qn}.}

Instead of calculating the spectral and statistical components 
of the self-energies, it is easier to calculate the
Wightman components
$\Sigma_\gtrless(x,y)\equiv \Sigma_F(x,y)\mp \frac{i}{2}
\Sigma_\rho(x,y)$. These can be identified with
the gain and loss terms, as mentioned earlier in this section.
After some simple but tedious algebra (see Appendix\,\ref{selfenergies} for more details), 
we obtain the self-energies corresponding to the diagrams in Fig.\,\ref{diagrams}: 
\begin{subequations}
	\label{Sigmagtrless}
	\begin{align}
		\label{Sigmab}
		\Sigma^{(b)}_\gtrless(x,y)&=
		-{\textstyle\frac12}\lambda^2 D_\gtrless (x,y)D_\gtrless (x,y)D_\lessgtr(y,x)\,,\\
		\label{Sigmac}
		\Sigma^{(c)}_\gtrless(x,y)&=-g_i g_j^*G^{ij}_\lessgtr(y,x)D_\lessgtr(y,x)\,,\\
		\label{Sigmad}
		\Sigma^{(d)}_\gtrless(x,y)&=-g_i g_j g_m^* g_n^*
		\textint \dg{v}\dg{u} \nonumber \\
		[\,D_R(y,v&)D_F(u,v)D_\lessgtr(u,x) 
		G^{ij}_R(y,u)G^{mn}_\gtrless(x,v)\nonumber\\
		+D_R(y,v&)D_A(u,v)D_\lessgtr(u,x) 
		G^{ij}_F(y,u)G^{mn}_\gtrless(x,v)\nonumber\\
		+D_F(y,v&)D_R(u,v)D_\lessgtr(u,x) 
		G^{ij}_R(y,u)G^{mn}_\gtrless(x,v)\nonumber\\
		+D_\lessgtr(y,v&)D_F(u,v)D_A(u,x) 
		G^{ij}_\lessgtr(y,u)G^{mn}_R(x,v)\nonumber\\
		+D_\lessgtr(y,v&)D_A(u,v)D_F(u,x) 
		G^{ij}_\lessgtr(y,u)G^{mn}_R(x,v)\nonumber\\
		+D_\lessgtr(y,v&)D_R(u,v)D_A(u,x) 
		G^{ij}_\lessgtr(y,u)G^{mn}_F(x,v)\nonumber\\
		+D_R (y,v&)D_\gtrless(u,v)D_A(u,x) 
		G^{ij}_\lessgtr(y,u)G^{mn}_\gtrless(x,v)\nonumber\\
		+D_\lessgtr(y,v&)D_\lessgtr(u,v)D_\lessgtr(u,x) 
		G^{ij}_R(y,u)G^{mn}_R(x,v)]\,.
	\end{align}
\end{subequations}
Diagrams $(b)$ and $(c)$ in Fig.\,\ref{diagrams} induce contributions 
$\Sigma^{(b)}$ and $\Sigma^{(c)}$ which contain only the Wightman two-point correlation functions
$D_{\gtrless}$ and $G^{ij}_{\gtrless}$. As we will show 
later, in the Boltzmann approximation these correspond to the on-shell 
initial and final states. To write the last term in compact form, we have 
introduced the retarded and advanced propagators, $D_R(x,y)\equiv \theta(x^0-y^0)D_\rho(x,y)$ 
and $D_A(x,y)\equiv -\theta(y^0-x^0)D_\rho(x,y)$, respectively.
The first six terms\footnote{The seventh term in \eqref{Sigmad}
describes the scattering process $\psi_i b\rightarrow \psi_i b$. 
This is clear from the fact that it contains one $D_\gtrless$
and two $G_\gtrless$ two-point functions, i.e.~one ``external'' 
complex scalar and two ``external'' real scalars. Similarly, the 
eighth term of Eq.\,\eqref{Sigmad} describes the scattering process
$bb\rightarrow \bar{b}\bar{b}$, because it contains three $D_\gtrless$ 
two-point functions, i.e.~three ``external'' complex scalars.} 
of $\Sigma^{(d)}$ describe the one-loop correction to the decay width. 
The combinations of the statistical, retarded and advanced propagators in 
Eq.\,\eqref{Sigmad} correspond to the three internal lines in the loop, 
whereas the $\gtrless$ components again correspond to the on-shell initial and final states.

Since we do not consider quartic interactions of the real scalar
fields, there are no local corrections to their masses. It is for 
this reason that Eqs. \eqref{Gequations} contain only the bare
masses $M_i^2$ of the fields. The first nonlocal term $\Pi^{(c)}$
describes the decay of the heavy scalar at tree-level:
\begin{subequations}
	\label{Pigtrless}
	\begin{align}
		\label{Pic}
		\Pi^{(c)ij}_\gtrless(x,y)&=-{\textstyle\frac12}g_i g^*_j D^2_\gtrless(x,y)
		-{\textstyle\frac12}g^*_i g_j D^2_\lessgtr(y,x)\,,\\
		\label{Pid}
		\Pi^{(d)ij}_\gtrless(x,y)&=-{\textstyle\frac12}g_i g_j g_m^* g_n^* \textint
		\dg v\dg u \nonumber\\
		[\,G^{mn}_F(v,&u)D_\gtrless(x,v)D_\gtrless(x,u)
		D_R(y,v)D_R(y,u)\nonumber\\
		+G^{mn}_R(v,&u)D_\gtrless(x,v)D_\gtrless(x,u)
		D_R(y,v)D_F(y,u)\nonumber\\
		+G^{mn}_A(v,&u)D_\gtrless(x,v)D_\gtrless(x,u)
		D_F(y,v)D_R(y,u)\nonumber\\
		+G^{mn}_F(v,&u)D_R(x,v)D_R(x,u)
		D_\lessgtr(y,v)D_\lessgtr(y,u)\nonumber\\
		+G^{mn}_R(v,&u)D_R(x,v)D_F(x,u)
		D_\lessgtr(y,v)D_\lessgtr(y,u)\nonumber\\
		+G^{mn}_A(v,&u)D_F(x,v)D_R(x,u)
		D_\lessgtr(y,v)D_\lessgtr(y,u)\nonumber\\
		+G^{mn}_\lessgtr(v,&u)D_\gtrless(x,v)D_R(x,u)
		D_R(y,v)D_\lessgtr(y,u)\nonumber\\
		+G^{mn}_\gtrless(v,&u)D_R(x,v)D_\gtrless(x,u)
		D_\lessgtr(y,v)D_R(y,u)]\nonumber\\
		&-{\textstyle\frac12}g^*_i g^*_j g_m g_n\textint
		\dg v\dg u \nonumber\\
		[\,
		G^{mn}_F(v,&u)D_\lessgtr(v,x)D_\lessgtr(u,x)
		D_A(v,y)D_A(u,y)\nonumber\\
		+\,\ldots\hspace{6mm}&\hspace{55mm} ]\,.
	\end{align} 
\end{subequations}
The first six terms\footnote{The seventh and eighth terms of Eq.\,\eqref{Pid} describe 
the scattering processes $\psi_i b\rightarrow \psi_j b$ and $b\bar b \rightarrow \psi_i\psi_j$.}
in each of the two square brackets of $\Pi^{(d)}$ 
describe the one-loop corrections to the scattering width. Their 
structure is very similar to that of the first six terms of Eq.\,\eqref{Sigmad} 
and the combinations of the statistical, retarded 
and advanced propagators again correspond to the three internal 
lines in the loop. 

The Kadanoff--Baym equations \eqref{Dequations} and \eqref{Gequations}
together with the expression for the self-energies \eqref{Sigmagtrless}
and \eqref{Pigtrless} form a closed system of integro-differential 
equations. Its solutions carry full information about the spectral 
and statistical properties of the system, including information about 
the generated asymmetry at each instant of time.

\subsection{Quantum-corrected Boltzmann equations}
Despite all advantages, the full Kadanoff--Baym equations  are 
relatively rarely used for the analysis of out-of-equilibrium 
processes, partially because of the complexity of the numerical
solution.\footnote{See e.g.~\cite{Danielewicz:1982ca,
Berges:2000ur,Berges:2001fi,Aarts:2001yn,Berges:2002wr,
Juchem:2003bi,Arrizabalaga:2005tf,Lindner:2005kv,Lindner:2007am}.}
In this section we derive kinetic equations starting from the above 
Kadanoff--Baym equations by applying a gradient expansion and a quasiparticle 
approximation \cite{Danielewicz:1982kk,Ivanov:1999tj,Knoll:2001jx,Blaizot:2001nr,
Weinstock:2005jw,Carrington:2004tm,FillionGourdeau:2006hi}.
The resulting quantum-corrected Boltzmann equations can be directly compared to the
canonical equations and are easier to solve numerically.

The quantum-corrected Boltzmann equations are applicable to weakly 
coupled systems of (quasi)particles that have a width which is small 
compared to their mass and that evolves slowly compared to the microscopic 
interaction time scales. We expect that these conditions are satisfied for thermal 
leptogenesis, where the deviations from equilibrium are moderate in general.
In thermal equilibrium the two-point correlation functions $D(x,y)$ 
and $G^{ij}(x,y)$ depend only on the relative coordinate, $s\equiv x-y$, and 
are independent of the center coordinate\footnote{The above definitions 
of the relative and center coordinates are valid only in Minkowski 
space-time. In a general space-time the center coordinate is defined 
as coordinates of the center ($\varsigma_X$) of the geodesic connecting 
$x$ and $y$, $X^\alpha \equiv \xi^\alpha(\varsigma_X)$, 
whereas the relative coordinate is proportional to the length of the geodesic 
between the two points, $s^\alpha\equiv u^\alpha(\varsigma_X)(\varsigma_x-\varsigma_y)$ 
\cite{PhysRevD.32.1871}.}, $X\equiv\frac12(x+y)$. Having these equilibrium considerations in 
mind, we trade the variables $x$ and $y$ for the new arguments $X$ and $s$: $D(x,y)
\rightarrow D(X,s)$. Out of equilibrium the two-point functions
depend on both the relative and center coordinate. If, however, 
the deviation from equilibrium is small, one can perform a gradient 
expansion of the correlation functions and the self-energies in the 
vicinity of  $X$ keeping only the leading  terms. This results 
in a system of equations describing the slow, $X$-dependent dynamics 
of the statistical properties of the system, see Appendix\,\ref{Complscal} 
for more details. Performing the gradient expansion, which is
the first step in the derivation of the Boltzmann equation, we replace the non-Markovian evolution equations by
a system of Markovian ones. Therefore, when truncating at first order in the gradient expansion, we neglect the memory effects.
The fast dynamics associated with the relative 
coordinate is responsible for the spectral properties of the system, 
which are conveniently described in the momentum representation. Performing 
the  Wigner transformation  (see Appendix\,\ref{Complscal}), we trade the 
relative coordinate $s$ for a coordinate $p$ in  momentum space: 
$D(X,s)\rightarrow D(X,p)$. To perform the Wigner transformation
we have to send the initial time to minus infinity which means that we neglect the effects of initial correlations.
The next step in the derivation of  Boltzmann 
equations is the  Kadanoff--Baym \textit{ansatz}. It relates the statistical 
propagator and the spectral function by a generalized fluctuation-dissipation
relation,
\begin{align}
	\label{KB_Ansatz}
	{D}_F(X,p)=\left[f(X,p)+{\textstyle\frac12}\right]
	{D}_{\rho}(X,p)\,,
\end{align}
where $f(X,p)$ is the one-particle distribution function. In the equilibrium limit, 
the Kubo-Martin-Schwinger condition \cite{Lebellac:2000bj,Landsman:1986uw}  ensures that $f(X,p)$ converges towards a Bose--Einstein
distribution function. The final step is the quasiparticle approximation, 
where one replaces the exact smooth spectral function by a Dirac $\delta$ function 
peaked on the mass-shell of the quasiparticles. The resulting Boltzmann-like equation, 
which describes the time evolution of the particle distribution function, reads
\begin{align}
	\label{DBlzmn_particles}
	[\,p^\alpha &{\cal D}_\alpha f(X,p)]{D}_{\rho}(X,p) \nonumber\\
	&={\textstyle\frac12}[{D}_{<}(X,p){\Sigma}_{>}(X,p)
	-{D}_{>}(X,p){\Sigma}_{<}(X,p)]\,.
\end{align}
The analogous equation for  antiparticles is given by
\begin{align}
	\label{DBlzmn_antiparticles}
	[\,p^\alpha &{\cal D}_\alpha \bar{f}(X,p)]{\bar D}_{\rho}(X,p)\nonumber\\
	&={\textstyle\frac12}[{\bar D}_{<}(X,p){\bar\Sigma}_{>}(X,p)
	-{\bar D}_{>}(X,p){\bar \Sigma}_{<}(X,p)]\,,
\end{align}
where ${\bar D}_{\gtrless}(X,p)\equiv {D}_{\lessgtr}(X,-p)$, 
${\bar \Sigma}_{\gtrless}(X,p)\equiv {\Sigma}_{\lessgtr}(X,-p)$.

Note again that by making the approximations which have led to 
\eqref{DBlzmn_particles} and \eqref{DBlzmn_antiparticles} we have 
neglected the memory effects and the effects of the initial correlations.
As a result, the quantum-corrected Boltzmann equations  are Markovian, 
i.e. local in time.

To obtain a closed system of quantum-corrected Boltzmann equations, we must also 
Wigner-transform the self-energies \eqref{Sigmagtrless} (see Appendix\,\ref{WignerTrafo} 
for more details). The latter encode the scattering and decay rates including
quantum nonequilibrium effects.
By employing the relations
\begin{align*}
 	 D_{<}(X,p) & =  f(X,p)  D_\rho(X,p) \;,\\
 	 D_{>}(X,p)  & =  [ 1 + f(X,p) ]  D_\rho(X,p) \;,
\end{align*}
which follow directly from Eq.\,(\ref{KB_Ansatz}), we
can then rewrite Eqs.\,(\ref{DBlzmn_particles},\ref{DBlzmn_antiparticles}) in a way
resembling the usual form of Boltzmann equations, including the correct
quantum statistical factors.

Within the 2PI three-loop approximation, we find that there are two physically 
distinct contributions to the self-energy. The first one, corresponding to $\Sigma^{(b)}$, 
describes \textit{CP}-conserving two body scatterings, $bb\rightarrow bb$, 
at tree-level:
\begin{align}
	\label{Sigma_scatt}
	\Sigma^{\times}_\gtrless(X,p)&=-{\textstyle\frac12}\lambda^2
	\textint d\Pi_{p_1}d\Pi_{p_2}d\Pi_{p_3}
	(2\pi)^4 \delta^g(p+p_1\nonumber\\
	-p_2&-p_3){D}_{\lessgtr}(X,p_1){D}_{\gtrless}(X,p_2){D}_{\gtrless}(X,p_3)\,,
\end{align}
where the invariant volume element in momentum space is defined by
\begin{align}
	d\Pi_p\equiv \frac1{\sqrt{-g}_X}\frac{d^4p}{(2\pi)^4}\,,
\end{align}
and  $\delta^g(p)\equiv \sqrt{-g}_X\, \delta(p)$ is the covariant 
generalization of the $\delta$ function \cite{Hohenegger:2008zk}.

The second contribution, given by the sum of $\Sigma^{(c)}$ and $\Sigma^{(d)}$,
describes decay processes $\psi_i\rightarrow bb$ and $\psi_i\rightarrow 
\bar{b} \bar{b}$ at tree- and one-loop level:
\begin{align}
	\label{Sigma_decay}
	\Sigma^{\sphericalangle}_\gtrless(X,&p)=-|g_i|^2 \textint d\Pi_{p_1}d\Pi_{p_2}
	(2\pi)^4 \delta^g(p_1-p_2-p)\nonumber\\
	&\times [1+\Delta_b^i(X,p_1,p_2)] {G}^{ii}_\gtrless(X,p_1){D}_{\lessgtr}(X,p_2)\,.
\end{align}
The newly introduced function $\Delta_b^i(X,p_1,p_2)$ takes into account 
the one-loop corrections to the decay width and is given by
\begin{align}
	\label{delta_def}
	\Delta_b^i(X,&\,p_1,p_2)=
	|g_j|^2 \left(\frac{g_i g^*_j}{g^*_i g_j}\right)
	\int d\Pi_{k_1}d\Pi_{k_2}d\Pi_{k_3} \nonumber\\
	&\times (2\pi)^4\delta^g(p_1+k_1+k_2) (2\pi)^4\delta^g(k_2-k_3+p_2)\nonumber\\
	&\times[D_A(X,k_1)D_F(X,k_2)G_A^{jj}(X,k_3)\nonumber\\
	&+\,D_A(X,k_1)D_R(X,k_2)G_F^{jj}(X,k_3)\nonumber\\
	&+\,D_F(X,k_1)D_A(X,k_2)G_A^{jj}(X,k_3)]+{\rm c.c.}
\end{align}
Proceeding in the same way, we derive quantum-corrected Boltzmann equations for the 
distribution functions of the real scalar fields, which is a two-by-two differential 
matrix equation. The off-diagonal components of the correlation functions 
are generated dynamically by the exchange of two complex scalars and 
are therefore of order  $g^2$. The one-loop vertex 
terms, which generate the \textit{CP}-violating parameter, are proportional 
to the fourth power of the coupling constant. Therefore the contribution 
of the off-diagonal terms to the vertex \textit{CP}-violating parameter is 
of the order of $g^6$. Here,  we limit ourselves to terms of at 
most fourth power in the coupling constant and therefore  we can neglect 
the off-diagonal terms in the corresponding matrix equation. The 
resulting equations coincide then with those derived in \cite{Hohenegger:2008zk}:
\begin{align}
	\label{GBlzmn}
	[\,p^\alpha & {\cal D}_\alpha  f_{\psi_i}(X,p)]{G}^{ii}_{\rho}(X,p)\nonumber\\
	&={\textstyle\frac12}[{G}^{ii}_{<}(X,p){\Pi}^{ii}_{>}(X,p)
	-{G}^{ii}_{>}(X,p){\Pi}^{ii}_{<}(X,p)]\,.
\end{align}
Note that we have in fact used this diagonal approximation in Eqs.\,\eqref{Sigma_decay} 
and \eqref{delta_def}. The Wigner-transform of the self-energy \eqref{Pigtrless} 
is given in the same approximation by
\begin{align}
	\label{Pi_decay}
	\Pi^{\sphericalangle\,ii}_\gtrless&(X,p)=-
	{\textstyle\frac12}|g_i|^2{\textstyle\int} d\Pi_{p_1}d\Pi_{p_2}
	(2\pi)^4 \delta^g(p_1+p_2-p)\nonumber\\
	&\times\{ [1+\Delta_\psi^i(X,p,p_2)]
	\,{D}_{\gtrless}(X,p_1){D}_{\gtrless}(X,p_2)\nonumber\\
	&+\,\,\,[1+\bar{\Delta}_\psi^i(X,p,p_2)] \, 
	{\bar D}_{\gtrless}(X,p_1){\bar D}_{\gtrless}(X,p_2)
	\}\,.
\end{align}
The second line of \eqref{Pi_decay} describes the process $\psi\rightarrow bb$.
The one-loop correction to this process is given by
\begin{align}
	\label{delta_psi}
	\Delta_\psi^i(X,\,&p,p_2)=
	|g_j|^2 \left(\frac{g_i g^*_j}{g^*_i g_j}\right)
	\int d\Pi_{k_1}d\Pi_{k_2}d\Pi_{k_3} \nonumber\\
	&\times (2\pi)^4\delta^g(p+k_1+k_2) (2\pi)^4\delta^g(k_2-k_3+p_2)\nonumber\\
	&\times[D_R(X,k_1)D_R(X,k_2)G_F^{jj}(X,k_3)\nonumber\\
	&+\,D_R(X,k_1)D_F(X,k_2)G_A^{jj}(X,k_3)\nonumber\\
	&+\,D_F(X,k_1)D_R(X,k_2)G_R^{jj}(X,k_3)]+{\rm c.c.}
\end{align}
The third line of \eqref{Pi_decay} describes  $\psi\rightarrow \bar{b}\bar{b}$ 
process and the corresponding one-loop contribution is related
to \eqref{delta_psi} by $\bar{\Delta}^i_\psi(X,p_1,p_2)\equiv 
\Delta^i_\psi(X,-p_1,-p_2)$.

A very important feature of the expressions for the self-energies,
Eqs.\,\eqref{Sigma_decay} and \eqref{Pi_decay}, is that the loop corrections
$\Delta^i_b$ and $\Delta^i_\psi$ appear as overall factors on the right-hand
sides of the corresponding quantum-corrected Boltzmann equations. Therefore, when using the
conventional approximations and integrating
Eqs.\,(\ref{DBlzmn_particles},\ref{DBlzmn_antiparticles}), we obtain the structure\footnote{For 
the numerical analysis we use the full Boltzmann equation, since the approximations 
required to obtain integrated rate equations are not appropriate within the toy model (see
Sec.\,\ref{Numerics}). Note, however, that the latter in fact contain \emph{averaged}
effective \CP-violating parameters (see Sec.\,\ref{CPviol} and Appendix\,\ref{app:ThermalAv}),
$\epsilon_i \rightarrow \langle\epsilon_i\rangle$.
}
\begin{align}
	\partial_t (n_b,n_{\bar{b}})\propto (1\pm \epsilon_i)(n_i-n^{eq}_i)\ldots\,.
\end{align}
To obtain an equivalent result in the canonical approach one explicitly needs to 
apply the RIS subtraction procedure. This means that, here, the structure of the
equations automatically ensures that no asymmetry is generated in thermal 
equilibrium. Stated differently, the Kadanoff--Baym formalism is \emph{free of the 
double-counting problem} and no need for RIS subtraction arises.

In the homogeneous and isotropic early universe the 
canonical Boltzmann equations conserve the linear combination 
$2n_{\psi_i}+n_b+n_{\bar{b}}$ of  particle numbers in a comoving volume.
However, the conservation of this  quantity is accidental, i.e.~not 
guaranteed by a symmetry of the underlying Lagrangian. It 
is not conserved by the full Kadanoff--Baym equations\footnote{
It is important to note that the 2PI approximation scheme guarantees that the 
total \textit{energy-momentum tensor} (including ``potential'' energy due 
to interactions in the system) is covariantly conserved
\cite{Ivanov:1998nv,Knoll:2001jx}. In other words, the nonconservation 
of the total number of particles does not violate the
fundamental conservation laws.} (see \cite{Lindner:2005kv} 
for another example). 
This is also true for the \textit{quantum-corrected Boltzmann equations} which we have derived from the Kadanoff--Baym 
equations. To see this one should add Eqs.\,\eqref{DBlzmn_particles} 
and \eqref{DBlzmn_antiparticles} as well as Eq.\,\eqref{GBlzmn} multiplied by two 
and use the explicit expressions for the self-energies \eqref{Sigma_decay} and 
\eqref{Pi_decay}. Although the expressions for the loop corrections \eqref{delta_def} 
and \eqref{delta_psi} are similar, they are not equal. This results in a
small time-dependence of the quantity
mentioned above (see also Appendix~\ref{app:NumericalDetails}).

The system of Boltzmann equations \eqref{DBlzmn_particles}, 
\eqref{DBlzmn_antiparticles} and \eqref{GBlzmn} together with the 
Wigner-transforms of the self-energies \eqref{Sigma_scatt}, 
\eqref{Sigma_decay} and \eqref{Pi_decay}  form a closed system of 
differential equations  that can be solved numerically. The 
solutions describe the phase-space distributions of quasiparticle
excitations at each instant of time $t=X^0$.

\section{\label{CPviol}\textit{CP}-violating parameter}
Comparing the Boltzmann equations for particles and antiparticles, 
\eqref{DBlzmn_particles} and \eqref{DBlzmn_antiparticles}, we see 
that the dynamical generation of the \textit{CP} asymmetry is only possible
if $\Sigma_\gtrless(X,p)\neq \bar{\Sigma}_\gtrless(X,p)$. Since 
$\bar{\Sigma}_\gtrless(X,p)\equiv \Sigma_\lessgtr(X,-p)$, in the 
diagonal approximation, this is equivalent to the requirement that
$\Delta^i_b(X,p_1,p_2)\neq \Delta^i_b(X,-p_1,-p_2)$. The \textit{CP}-violating 
parameter can then be defined as 
\begin{align}
	\hskip -1mm
	\epsilon_i(X,p_1,p_2)\equiv{\textstyle\frac12}[\Delta^i_b(X,p_1,p_2)- 
	\Delta^i_b(X,-p_1,-p_2)].
\end{align}
Using properties of the Wigner transforms of the statistical, 
retarded and advanced propagators, we find that in a medium that
is (approximately) baryosymmetric\footnote{We will see later  
that the \textit{CP}-violating parameter, defined in this way, is different for 
particles and antiparticles if the corresponding distribution functions are different. 
Since the expected asymmetry is small, this is only a second order effect 
and can be neglected in the present work. The condition of almost zero asymmetry is certainly 
satisfied if the \textit{CP}-violating parameter is calculated in vacuum, 
as it is the case in the canonical approach.}
it is given by
\begin{align}
	\label{epsilon0}
	\epsilon&_i(X,p_1,p_2)=
	|g_j|^2 {\rm Im}\left(\frac{g_i g^*_j}{g^*_i g_j}\right)
	\int d\Pi_{k_1}d\Pi_{k_2}d\Pi_{k_3} \nonumber\\
	&\times (2\pi)^4\delta^g(p_1+k_1+k_2) (2\pi)^4\delta^g(k_2-k_3+p_2)\nonumber\\
	&\times[D_\rho(X,k_1)D_F(X,k_2)G_h^{jj}(X,k_3)+\{k_1\leftrightarrow k_2\}\nonumber\\
	&\,+D_h(X,k_1)D_F(X,k_2)G_\rho^{jj}(X,k_3)+\{k_1\leftrightarrow k_2\}\nonumber\\
	&\,+D_\rho(X,k_1)D_h(X,k_2)G_F^{jj}(X,k_3)-\{k_1\leftrightarrow k_2\}]\,,
\end{align}
where $D_h(X,p)\equiv \Re D_R(X,p)=\Re D_A(X,p)$ and $G_h$ is 
defined analogously. The quasiparticle approximation 
together with the Kadanoff--Baym ansatz enforces the spectral functions 
and the statistical propagators to be on  mass-shell. On the contrary, the real 
parts of the retarded propagators, $D_h$ and $G_h$, vanish on-shell 
(see Appendix\,\ref{Complscal} for more detail). 
\begin{figure}[!ht]
	\begin{center}
	\includegraphics[width=0.45\textwidth]{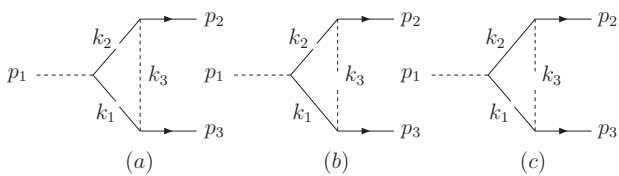}
	\end{center}
\caption{\label{cpcontributions}Graphical representation of the terms in 
Eq.\,(\ref{epsilon0}).}
\end{figure} 
In other words,  in the processes contributing 
to the \textit{CP}-violating parameter, two of the intermediate lines are on-shell
and one line is off-shell; this is shown in Fig.\,\ref{cpcontributions}.
It is also interesting to note that in each term on the right-hand side 
of \eqref{epsilon0} only one of the internal lines is ``thermal'', i.e.~explicitly 
depends on the one-particle distribution function. 
In other words the medium corrections are only \textit{linear} in the 
particle number densities.
One could come to this conclusion even without explicitly calculating the 
\textit{CP}-violating parameter. The self-energy \eqref{Sigmad} is  a 
product of five two-point correlation functions. As far as the decay is 
concerned, two of them describe on-shell ``external'' states. The two 
integrals over the space-time coordinates, $u$ and $v$, turn two other 
functions  into  retarded and (or) advanced propagators, which do not 
explicitly depend on the particle number densities. The remaining function 
turns out to be given by the corresponding statistical propagator and 
explicitly depends on the particle number density.

Typically, one is interested only in the total generated asymmetry and 
solves  rate equations for the total particle number densities.
They are obtained by integrating the left- and right-hand sides
of the Boltzmann equations \eqref{DBlzmn_particles} and \eqref{DBlzmn_antiparticles} 
over phase space. In the corresponding gain and loss terms,
\begin{align}
	\label{gainloss}
	\textint d\Pi_{p} d\Pi_{p_1}d\Pi&_{p_2}
	(2\pi)^4 \delta^g(p_1-p_2-p)[1\pm\epsilon_i(X,p_1,p_2)]\nonumber\\
	&\times  {G}^{ii}_\gtrless(X,p_1){D}_{\lessgtr}(X,p_2) 
	{D}_{\lessgtr}(X,p)\,,
\end{align}
we can perform the transformation $p\leftrightarrow p_2$ and take a sum
of the initial and final expressions, so that in Eq.\,\eqref{gainloss}:
\begin{align}
	\label{epsilonint}
	\epsilon_i(X,p_1,p_2)\rightarrow {\textstyle\frac12}
	[\epsilon_i(X,p_1,p_2)+\epsilon_i(X,p_1,p)]\,.
\end{align}
In $\epsilon_i(X,p_1,p)$, we transform the variables $k_1\leftrightarrow k_2$ and
$k_3\rightarrow -k_3$ in addition.
The spectral function
of a real scalar field is antisymmetric: $G^{ii}_\rho(X,-k_1)=
-G^{ii}_\rho(X,k_1)$. Collecting all the terms, we find that due to the antisymmetry 
only the first two terms (the third line) in \eqref{epsilon0} contribute to the right-hand 
side of \eqref{epsilonint}, whereas the other terms cancel out. In other words, only the 
first two terms of \eqref{epsilon0} contribute to the total \textit{CP}-asymmetry, whereas
the other four terms do not. An explicit calculation shows that (at least in the homogeneous 
and isotropic universe) these four terms also do not contribute to the gain and loss terms.
Diagrammatically this means that only decays followed by a scattering 
contribute to the \textit{CP}-violating parameter, see Fig.\,\ref{cpcontributions}a.

As we have already mentioned, it is important that the \textit{CP}-violating
parameters $\epsilon_i$ are identical\footnote{It is important to note
that the transformation properties under $C$, $P$ and $T$ of the 
self-energies obtained from the Schwinger--Keldysh/Kadanoff--Baym formalism
cannot be identified with those of the S-Matrix elements appearing in the canonical in-out 
formalism.} for the gain and the loss terms: this means that the structure of the
quantum-corrected Boltzmann equation automatically ensures that
the asymmetry vanishes in thermal equilibrium. Let us also note 
that there is a clear distinction between the initial, final and 
 intermediate states: the former ones are described by 
the Wightman functions $D_{\gtrless}$ and $G_{\gtrless}$, 
whereas the latter ones are described by the
retarded, advanced and (or) statistical components of the
two-point functions.

Applying the quasiparticle approximation and the Kadanoff--Baym 
ansatz  in Eq.\,(\ref{epsilon0}), we obtain the following expression for the \textit{CP}-violating 
parameter which is one of our central results:
\begin{align}
	\label{epsilon_expl}
	\epsilon_i(p_1,p_2)=&-\frac1{8\pi}\frac{|g_j|^2}{M_i^2}{\rm Im}\left(\frac{g_ig^*_j}{g^*_ig_j}\right)\nonumber\\
	&\times\int\frac{d\Omega}{4\pi}\frac{1+\bar{f}(E_{k_1})+\bar{f}(E_{k_2})}{
	M_j^2/M_i^2+\frac12(1+\cos\theta)}\,,
\end{align}
where $E_{k_{1,2}}$ are the energies of the intermediate toy bary\-ons as a function
of $p_1$, $p_2$ and the angle variables, and we have 
omitted the time-space coordinate $X$ to shorten the notation.
The \textit{CP}-violating 
parameter is  a sum of  vacuum and medium contributions. 
Integrating the vacuum contribution over the solid angle, we obtain 
the standard expression for the \textit{CP}-violating parameter, see 
Eq.\,\eqref{epsilonvac}. The thermal contributions are proportional to 
the one-particle distribution function, which is  positive. 
Hence, for scalars the \textit{CP}-violating parameter is always 
\textit{enhanced} by the medium effects. 

Because of the fact that the intermediate toy baryons propagate with respect to the
rest frame of the thermal bath, the \textit{CP}-violating parameter 
depends in each individual decay  on the phase-space distribution 
of the decaying particle and the decay products. Using results of 
Appendix\,\ref{Kinematics}, we obtain for the energy of the intermediate 
complex scalars:
\begin{align}
	\label{IntermedEnergy}
	\hskip -1mm
		E_{k_{1,2}}={\textstyle\frac12}[E_1+|\vec{p}_1|(\sin\theta\cos\varphi\cos\delta' \mp \cos\theta\sin\delta')],
\end{align}
where $\theta$ and $\varphi$ are elements of the solid angle $\Omega$
and the angle $\delta'$ depends on momenta of the initial and final states:
$\sin\delta'=(|\vec{p}_3|-|\vec{p}_2|)/|\vec{p}_1|$. 

In the limit of almost equal one-particle distribution functions of 
particles and antiparticles, $f$ and $\bar f$, the \textit{CP}-violating
parts of $\Delta^i_\psi$ do not contribute to the Boltzmann equations
for the real scalars, just as in the canonical 
approach. For this reason, we do not consider it here.

In order to estimate the size of the medium corrections, we consider the hierarchical
limit of the heavy scalar mass spectrum, $M_1 \ll M_2$. As in standard leptogenesis, we
assume that the asymmetry is predominantly generated by the decay of the lighter
scalar. By expanding Eq.\,(\ref{epsilon_expl}) in $M_1^2/M_2^2$, we obtain a simplified
expression for the  relevant \textit{CP}-violating parameter $\epsilon_1$,
\begin{equation}\label{epsilon_hierarchical}
	\epsilon_1(p_1,p_2) = \epsilon_1^{\it vac} \left[ 1 + \textstyle{\int\frac{d\Omega}{4\pi}} 
	\left\{ \bar{f}(E_{k_1}) + \bar{f}(E_{k_2}) \right\}\right] \; ,
\end{equation}
where $\epsilon_1^{\it vac}$ is the \textit{CP}-violating parameter in 
vacuum given in Eq.\,(\ref{epsilonvac}). Exploiting the $k_1\leftrightarrow k_2$
symmetry and integrating over the full solid angle we find 
that the asymmetry depends  on the absolute value  of $\vec{p}_1$ only. That is
$\epsilon_1(p_1,p_2)=\epsilon_1(|\vec{p_1}|)$, where
\begin{align}\label{epsilon hierarchical massless}
	\epsilon_1(|\vec{p}|)=&\epsilon_1^{vac}\left[1+{  \frac{2}{r\left|\vec{p}\right|}} 
	\int_{E_{min}/2}^{E_{max}/2} \bar{f}(E) dE\right]\,,
\end{align}
and $E_{\it max}=E_1+r|\vec{p}|$ and $E_{\it min}=E_1-r|\vec{p}|$ are the largest
and smallest kinematically allowed energies of the light scalars produced in the decay
$\psi_1\rightarrow bb$. Here $E_1=(M_1^2+\vec{p}^2)^\frac12$ and $|\vec{p}|$ denote the
energy and momentum of $\psi_1$ in the rest frame of the medium, respectively.
Furthermore, we have also included a nonzero ``baryon'' mass $m$ for completeness,
which enters via the parameter $r\equiv(1-4m^2/M_1^2)^\frac12$. 
\begin{figure}[h!]
	\includegraphics[width=\columnwidth]{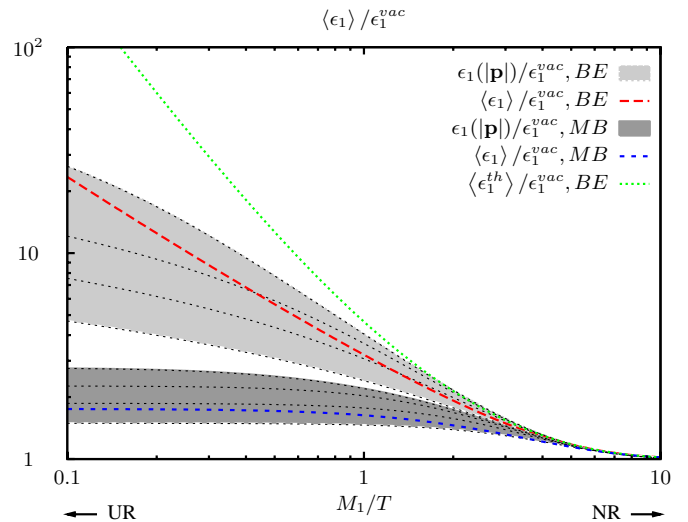}
	\caption{\label{thermalEpsilonAveraged}%
		Effective \CP-violating parameter $\epsilon_1(|\vec{p}|)$ in medium obtained from the
		Kadanoff-Baym formalism. The shaded areas correspond to the range $0.25 \le |\vec{p}|/T \le 4$
		of momenta $|\vec{p}|$ of the decaying particle $\psi_1\rightarrow bb/\bar b\bar b$ with
		respect to the rest frame of the medium. Here we assumed a thermal Bose-Einstein (BE) and
		Maxwell-Boltzmann (MB) distribution for $b$/$\bar b$ with vanishing chemical potential for
		illustration. In the low-temperature limit (NR), the vacuum value is approached.
		In the high-temperature limit (UR), the \CP-violating parameter is enhanced for bosons.
		We also show the thermally averaged \CP-violating parameter $\langle\epsilon_1\rangle$
		for the BE (red long-dashed line) and MB (blue dashed line) cases, as well as the
		result that would be obtained in thermal field theory (green dotted line).
	}
\end{figure}

The medium correction depends on the one-particle distribution function of the light
scalars  and the masses of the particles. As expected they vanish if $\bar{f}\equiv 0$.
We emphasize that the upper expression is valid even if the light
scalars were out of equilibrium. Nevertheless, since we expect the light scalars to be
close to kinetic equilibrium at all times, we insert a Bose--Einstein distribution
function. For comparison we also consider the case of a Maxwell--Boltzmann
distribution. Then  we obtain
\begin{equation}\label{epsilon_hierarchical_BE}
	\frac{ \epsilon_1(|\vec{p}|) }{ \epsilon_1^{\it vac} } \, = \, 1 + \frac{2T}{r|\vec{p}|} \times \left\{
	\begin{array}{ll}
		\displaystyle \ln \left( \frac{ 1 - \exp\left( - \frac{E_{\it max} -2\mu }{ 2T } \right) }
		{ 1 - \exp\left( - \frac{E_{\it min} - 2\mu }{ 2T } \right) } \right) &  \mbox{BE} \; , \\[4.5ex]
		\displaystyle e^{ - \frac{E_{\it min} -2\mu }{ 2T } } - e^{ - \frac{E_{\it max} -2\mu }{ 2T } } &  \mbox{MB} \; .
	\end{array}\right.
\end{equation}
The resulting expression depends on time $t=X_0$ via temperature $T$
and chemical potential $\mu$ of the  toy baryons.
For the rest of this section we assume $|\mu|\ll T$, as in realistic scenarios of leptogenesis, for the
purpose of illustration\footnote{In realistic scenarios, the leptons and Higgs fields are in equilibrium with
gauge bosons such that $\mu = - \bar \mu$. The smallness of the asymmetry then ensures the
smallness of the chemical potentials, $|\mu|/T = |\bar \mu|/T\sim 10^{-10}$. Within the toy model,  due to 
the absence of gauge interactions, it is possible to have $|\mu| \simeq |\bar\mu| \sim T$
while the asymmetry remains small. It turns out that this is even necessary to obtain consistent
numerical solutions within the present scenario, see Sec.~\ref{Numerics}.
}.
The temperature and momentum dependence of the medium correction in the range of typical momenta 
$|\vec{p}| \sim T$ is shown in the shaded areas in Fig.~\ref{thermalEpsilonAveraged} for the BE and 
MB cases, respectively. It is instructive to consider the nonrelativistic (NR) regime ($T,|\vec{p}| \ll M_1$) and
the ultrarelativistic (UR) regime ($T \gtrsim |\vec{p}| \gg M_1$).
In the nonrelativistic (NR) limit, the BE and MB cases coincide, and the
medium correction is exponentially suppressed,
\begin{equation}
 \epsilon_1(|\vec{p}|)/\epsilon_1^{\it vac} \rightarrow 1 + 2 \; \exp\left(-\frac{M_1}{2T}\right) \;.
\end{equation}
Furthermore, it is independent of the momentum $|\vec{p}|$.
In the UR limit, the medium correction for the BE and MB cases behaves
quite differently: In the MB case the medium correction saturates at $\epsilon_1/\epsilon_1^{\it vac}\lesssim 3$.
In the BE case, it
is logarithmically enhanced\footnote{Note that the logarithmic enhancement at high energies 
is cut-off for extremely high energies ($|\vec{p}|\gg M_1^2/m$) in the UV due to the second 
summand of the denominator inside the logarithm. Since we assume that $m \ll M_1/10$, we
can neglect this term in the relevant temperature range $M_1/T>0.1$.
}
(see Fig.\,\ref{thermalEpsilonAveraged}),
\begin{equation}\label{epsilon_hierarchical_BE_Limits}
	\epsilon_1(|\vec{p}|)/\epsilon_1^{\it vac} \rightarrow 1 + \frac{2T}{r|\vec{p}|} \ln\left( \frac{4T|\vec{p}|}
	{M_1^2 + \frac{8\vec{p}^2m^2}{M_1^2(1+r)} } \right) \; .
\end{equation}
This effect is due to Bose  enhancement.
Thus, we find that the quantum statistics is important for the medium correction.
In the following section, we will see that the
logarithmic enhancement at high energies is also suppressed by the inclusion of
sizable negative chemical potentials (which is necessary within the toy model, see below).
In Fig.\,\ref{thermalEpsilonAveraged}, we also show the \CP-violating parameter $\langle\epsilon_1\rangle$ 
obtained from averaging Eq.\,(\ref{epsilon_hierarchical_BE}) over the momentum $|\vec{p}|$
(see Appendix\,\ref{app:ThermalAv}). As expected, $\langle\epsilon_1\rangle \sim \epsilon_1(|\vec{p}|\sim T)$.

Before discussing the impact of the medium correction quantitatively, we would
like to comment on the relation between the Kadanoff-Baym (top-down) and the canonical (bottom-up)
approach. As has been mentioned before,  \emph{in vacuum}, the top-down result Eq.\,(\ref{epsilon_expl})
for the \textit{CP}-violating parameter coincides with the canonical result
Eq.\,(\ref{epsilonvac}). Nevertheless, we emphasize again that the structure
of the Boltzmann equations differs between the two approaches, i.e. the former are
free of the double-counting problem. Furthermore, it is also important to note that
the size of the medium correction differs between the top-down and the
bottom-up approach. Within the latter, the medium corrections have been discussed
by replacing $\epsilon_i^{\it vac} \rightarrow \epsilon_i^{\it th}$ in the
canonical Boltzmann equations. Hereby $\epsilon_i^{\it th}$
involves the vertex loop calculated within thermal field theory
(see e.g.~\cite{Giudice:2003jh,Covi:1997dr}). For the toy model, $\epsilon_i^{\it th}$ is given in
Eq.\,(\ref{epsilon_th}) of Appendix\,\ref{cpclassic}. It involves an
additional term compared to the top-down result~(\ref{epsilon_expl}), which is quadratic in the
particle distribution function. In Fig.\,\ref{thermalEpsilonAveraged}
we show that the medium correction would be significantly over-estimated
in the canonical thermal field theory approach within the toy model.

For realistic models of leptogenesis, the vertex loop contains scalar and
fermionic lines in general. In contrast to the scalars, the latter tend to \emph{decrease}
the size of the \CP-violating parameter. Therefore the results shown
here can only be used indirectly to make statements for phenomenology. For the standard
scenario of thermal leptogenesis with hierarchical right-handed neutrino spectrum,
it has been observed in~\cite{Covi:1997dr} that, within thermal field theory, the effects
of Bose enhancement and Pauli blocking tend to cancel each other. However, since the
medium corrections differ within the Kadanoff-Baym formalism,
this cancellation may no longer occur. Therefore one might expect that the
medium correction is underestimated in this case, contrary to the situation encountered
within the toy model discussed here.

Note that we have neglected the effect of thermal masses here for simplicity.
Their impact on leptogenesis has been studied e.g.~in \cite{Giudice:2003jh} 
within the framework of thermal field theory. For the toy model considered 
here, it is consistent to neglect thermal masses due to the absence of gauge 
interactions. We stress, however, that it is also possible to include 
time-dependent effective masses systematically within the Kadanoff-Baym 
formalism. One of the effects caused by the thermal  masses of the 
complex field is the cutoff of the logarithmic enhancement of the \CP-violating 
parameter at high energies. In addition, in an asymmetric medium the effective
masses of the particles and antiparticles are not equal, which leads to 
an effective \CP~violation. This effect is studied in \cite{Garny:2009c}. 
The medium corrections to the masses of the heavy real fields can be important
in the case of the degenerate mass spectrum, which we investigate in 
\cite{Garny:2009qn}.

Since the quantum statistic is important for the \CP-violating parameter in medium,
we  also have to include the quantum statistical terms in the gain and loss terms of the 
Boltzmann equations for consistency. This will be discussed in the following section.

\section{\label{Numerics}Numerical results}
To obtain the three Boltzmann equations for $f$, $\bar{f}$ and $f_{\psi_1}$, we integrate 
each of  Eqs.\,(\ref{DBlzmn_particles}), (\ref{DBlzmn_antiparticles}) and (\ref{GBlzmn}) 
over $p_0$ (left- and right-hand side) and choose the positive energy solution \cite{Hohenegger:2008zk}.
In agreement with the cosmological principle, we solve the system of Boltzmann equations 
with spatially homogeneous and momentum isotropic distribution functions 
in (flat and radiation dominated) Friedman--Robertson--Walker (FRW) space-time. 
In this case the left-hand side of the Boltzmann equation is given by
\begin{align}
	\lorig [f](\momabs)\equiv&p^\alpha {\cal D}_\alpha f(\momabs)\nonumber\\
	=&p^0\left(\frac{\partial }{\partial t}-\momabs\hubblerate\frac{\partial }{\partial \momabs}\right)f(\momabs)\,,
\end{align}
where  $\hubblerate\equiv {\dot{a}}/{a}$ is the Hubble parameter. Since in the quasiparticle 
approximation the spectral functions are proportional to $\delta(p^2-m^2)$, the
integration over the time components of the quasiparticle's four-momenta
can be performed trivially in each of the corresponding self-energies. After 
the integration the volume element $d\Pi_p$ is replaced by $d\Pi^3_{p}\equiv 
d^3p/(2\pi)^3/2E$, where $E$ is energy of the on-shell quasiparticle.

	With these modifications, the network of quantum-corrected Boltzmann equations takes the form:
\begin{widetext}
	\begin{subequations}
		\label{QuantumBoltzmannEquations}
		\begin{align}
			\label{boltzmann equation b}
			\lorig [f](\momkabs)&=\carrayorig{bb\leftrightarrow bb}{\momkabs}{f}+
			\carrayorig{b\bar{b}\leftrightarrow b\bar{b}}{\momkabs}{f,\bar{f}}+ 
			\carrayorig{bb\leftrightarrow \psi_1}{\momkabs}{f,f_{\psi_1}}\,,\\
			\label{boltzmann equation bbar}
			\lorig [\bar{f}](\momkabs)&=\carrayorig{\bar{b}\bar{b}\leftrightarrow 
			\bar{b}\bar{b}}{\momkabs}{\bar{f}}+\carrayorig{\bar{b}b\leftrightarrow 
			\bar{b}b}{\momkabs}{\bar{f},f}+\carrayorig{\bar{b}\bar{b}\leftrightarrow 
			\psi_1}{\momkabs}{\bar{f},f_{\psi_1}}\,,\\
			\label{boltzmann equation psi}
			\lorig [f_{\psi_1}](\momkabs)&=\carrayorig{\psi_1\leftrightarrow {b}{b}}{\momkabs}{f_{\psi_1},f}
			+\carrayorig{\psi_1\leftrightarrow \bar{b}\bar{b}}{\momkabs}{f_{\psi_1},\bar{f}}\,,
		\end{align}
	\end{subequations}
	where the different collision terms for the transition between two states $i$ and $f$ are denoted by $\carrayorig{i\leftrightarrow f}{}{}$. Note
        that, due to the isotropy of the FRW universe, the collision terms also depend only on the absolute value of the momenta. For the $2-2$ scattering
        processes 
	in (\ref{boltzmann equation b}) we find
	\begin{subequations}
		\label{2bodyscattering}
		\begin{align}\label{collsion term bb-bb}
			\carrayorig{bb\leftrightarrow bb}{\momkabs}{f}& =
			{\textstyle\frac{1}{2}}\textint d\Pi^3_{\momp} d\Pi^3_{\momq} d\Pi^3_{\momr}
			(2\pi)^4 \delta^{(4)}(p+{\momp}-{\momq}-{\momr})\nonumber\\
			&\times {\textstyle\frac12}\lambda^2 \{[\qstat{f({\momkabs})}][\qstat{f({\mompabs})}]f({\momqabs})f({\momrabs})
			-f({\momkabs})f({\mompabs})[\qstat{f({\momqabs})}][\qstat{f({\momrabs})}]\}\,,\\
			\label{collsion term bbbar-bbbar}
			\carrayorig{b\bar{b}\leftrightarrow b\bar{b}}{\momkabs}{f,\bar{f}}&= 
			{\textstyle\frac{1}{2}}\textint d\Pi^3_{\momp} d\Pi^3_{\momq} d\Pi^3_{\momr}
			(2\pi)^4 \delta^{(4)}(p+{\momp}-{\momq}-{\momr})\nonumber\\
			&\times \lambda^2\{[\qstat{f({\momkabs})}][\qstat{\bar{f}({\mompabs})}]\bar{f}({\momqabs})f({\momrabs})-
			f({\momkabs})\bar{f}({\mompabs})[\qstat{\bar{f}({\momqabs})}][\qstat{f({\momrabs})}]\}\,.
		\end{align}
	\end{subequations}
	The corresponding terms in the equation for $\bar{b}$ can be obtained by replacing $f$ 
	with $\bar{f}$ in (\ref{collsion term bb-bb}) and (\ref{collsion term bbbar-bbbar}). If the generated 
	asymmetry is small, as we assume here, then $f\approx \bar f$. In this case the CP-violating 
	contributions to the right-hand side of \eqref{boltzmann equation psi} cancel out and we obtain:
	\begin{align}
		\label{collision term psi-bb and psi-bbar bbar}
		\carrayorig{\psi_1\leftrightarrow bb}{\momkabs}{f_{\psi_1},{f}}+
		\carrayorig{\psi_1\leftrightarrow \bar{b}\bar{b}}{\momkabs}{f_{\psi_1},\bar{f}} 
		\simeq{\textstyle\frac{1}{2}}\textint d\Pi^3_{\momp} d\Pi^3_{\momq} (2\pi)^4 & \delta^{(4)}(p-{\momp}-{\momq})\nonumber\\
		\times{\textstyle\frac12}|g_1|^2 (\{[\qstat{f_{\psi_1}({\momkabs})}]f({\mompabs})f({\momqabs})
		&-f_{\psi_1}({\momkabs})[\qstat{f({\mompabs})}][\qstat{f({\momqabs})}]\}\nonumber\\
		+\{[\qstat{f_{\psi_1}({\momkabs})}]\bar{f}({\mompabs})\bar{f}({\momqabs})
		&-f_{\psi_1}({\momkabs})[\qstat{\bar{f}({\mompabs})}][\qstat{\bar{f}({\momqabs})}]\})\,.
	\end{align}
	The collision terms for the (inverse) decay 
	of the heavy particle into a $bb$ or $\bar b \bar b$ pair explicitly contain the 
	\CP-violating parameter $\epsilon$ defined in Eq.\,\eqref{epsilon hierarchical massless}:
	\begin{subequations}
		\label{collision_term_bbbarbbarb-psi}
		\begin{align}
			\label{collision term bb-psi}
			\carrayorig{bb\leftrightarrow \psi_1}{\momkabs}{f,f_{\psi_1}}&=
			{\textstyle\frac{1}{2}}\textint d\Pi^3_{\momp} d\Pi^3_{\momq} (2\pi)^4 \delta^{(4)}({\momp}-p-{\momq})\nonumber\\
			&\times
			|g_1|^2[1+\epsilon_1(\left|\vec{p}_2\right|)]
			\{[\qstat{f({\momkabs})}][\qstat{f({\momqabs})}]f_{\psi_1}({\mompabs}) -f({\momkabs})f({\momqabs})[\qstat{f_{\psi_1}({\mompabs})}]\}
			\\
			\label{collision term bbar bbar-psi}
			\carrayorig{\bar{b}\bar{b}\leftrightarrow \psi_1}{\momkabs}{\bar{f},f_{\psi_1}}&=
			{\textstyle\frac{1}{2}}\textint d\Pi^3_{\momp} d\Pi^3_{\momq} (2\pi)^4 \delta^{(4)}({\momp}-p-{\momq})\nonumber\\
			&\times
			|g_1|^2[1-\epsilon_1(\left|\vec{p}_2\right|)]
			\{[\qstat{\bar{f}({\momkabs})}][\qstat{\bar{f}({\momqabs})}]f_{\psi_1}({\mompabs})
			-\bar{f}({\momkabs})\bar{f}({\momqabs})[\qstat{f_{\psi_1}({\mompabs})}]\}\,.
		\end{align}
	\end{subequations}
\end{widetext}
The factors $1/2$ associated with the couplings in (\ref{collsion term bb-bb})
and \eqref{collision term psi-bb and psi-bbar bbar} correctly account for the symmetrization of 
collision integrals which include integration over the momenta of two identical particles in the 
initial/final state. They have been consistently obtained in the derivation from the Kadanoff--Baym equations. 
We would like to stress again  that the structure of \eqref{collision_term_bbbarbbarb-psi} 
differs from the usual structure obtained in the conventional bottom-up approach. 
In particular, we do not need to include collision terms for the processes $bb\leftrightarrow \bar{b}\bar{b}$ 
(not even the RIS part of it) because our collision terms for the processes $bb\leftrightarrow\psi_1$ 
and $\bar{b}\bar{b}\leftrightarrow\psi_1$ do not suffer from the  generation of an asymmetry in equilibrium. 
The network of Boltzmann equations \eqref{QuantumBoltzmannEquations} should be understood 
in the generalized sense: the ``amplitudes'' differ from the usual perturbative matrix elements and 
do not have their symmetry properties.

To stay consistent in our model we also keep the quantum statistical factors for bosons. 
The implications of this new structure for a phenomenological theory of leptogenesis will be 
discussed elsewhere. To study the effect of the quantum corrections, we compare the results 
obtained by integrating the network of Boltzmann equations with quantum-corrected 
$\epsilon_1(\left|\vec{p}\right|)$ to those which are obtained after replacing $\epsilon_1(\left|\vec{p}\right|)$ 
with $\epsilon_1^{vac}$. This means that we keep here the new structure of the Boltzmann equations 
and study corrections which arise from the quantum-corrected $\epsilon_1$ only.
In the vacuum limit the structure of Eq.\,\eqref{QuantumBoltzmannEquations} corresponds 
to the one which has been assumed implicitly in \cite{HahnWoernle:2009qn} if the quantum 
statistical factors and the symmetrization factor are replaced accordingly.

In both cases we start at sufficiently high temperatures so that all species, including $\psi_1$ with 
mass $M_1=\Cmasspsi\,\mbox{GeV}$, have relativistic initial abundances which
corresponds to the most frequently discussed case.\footnote{Another scenario,
which is frequently discussed in the literature, is that the Majorana neutrinos
could have zero initial abundance. In this case we would expect the differences
in the time evolution of the asymmetry to be larger in general. However, this
can have an effect on the final asymmetry only if the asymmetry, produced in the
intermediate step of thermalization, is not washed out again before the Majorana
neutrinos decouple, i.e.~for small washout factor $\kappa\lesssim 1$.} Because
of the presence of the statistical factors we need to start with sufficiently
negative chemical potentials as to avoid Bose--Einstein condensation of the
different species during their evolution.\footnote{In this regime it would not be
appropriate to describe the system by conventional Boltzmann kinetic equations.
Since we are interested in scenarios that are qualitatively similar to realistic
models of leptogenesis here, we do not consider this case.}
We choose them such that they are related by $\mu_{\psi_1}=2\mu_b=2\mu_{\bar{b}}$, 
i.e.~the system is in chemical equilibrium. 

The coupling $\lambda$ can be adjusted such that the rates of the $2-2$ interactions 
\eqref{2bodyscattering} are much larger than those of the decays and inverse decays 
\eqref{collision term psi-bb and psi-bbar bbar}-\eqref{collision_term_bbbarbbarb-psi} 
at all times\footnote{As we will show in Appendix\,\ref{app:NumericalDetails} 
in this case there is no need to compute the collision integrals for $2-2$ scattering 
explicitly and we can use perturbative values for $\lambda$ for most of the relevant 
range of $\left|g\right|^2$ assuming that it is sufficient to demand that the rate for 
$b\bar{b}\leftrightarrow b\bar{b}$ is at least $10^3$ times larger than that of $bb\leftrightarrow{\psi_1}$.}. 
This keeps $b$ and $\bar{b}$ close to kinetic equilibrium, just as Higgs particles and 
leptons are kept in equilibrium by rapid gauge interactions in the standard scenario. 
The distribution functions for these species are therefore given by their equilibrium 
form throughout the entire evolution. This means that they can be described in terms of 
four parameters $\mu_b$, $T_b$ and ${\mu}_{\bar{b}}$, ${T}_{\bar{b}}$. The interactions \eqref{2bodyscattering} 
enforce the relation ${T}_{\bar{b}}=T_b$ between the parameters. Therefore, it is sufficient to study 
the evolution of $f$ and $\bar{f}$ in terms of the remaining three parameters 
(see Appendix\,\ref{app:NumericalDetails}). The evolution of ${\psi_1}$, however, 
is studied in terms of the complete distribution function discretized on a grid with $\Cdim$ momentum modes.
 Our computation, therefore, includes classical nonequilibrium effects in the decay of $\psi_1$.
 Such effects have been studied recently in \cite{Basbol:2007,HahnWoernle:2009qn,Garayoa:2009my}. 
All integrals are evaluated numerically including all quantum statistical factors for stimulated emission.

We define the generated ``baryon'' asymmetry as
\begin{equation}
	\eta(M_1/T) = \frac{ n_b(M_1/T) -n_{\bar{b}}(M_1/T) }{ s(M_1/T) } \;.
\end{equation}
Here $n_{b}$ and $n_{\bar{b}}$, the number densities of species $b$ and $\bar{b}$
respectively (compare \eqref{number density and energy density}),  are computed in the presence 
of the quantum-corrected $\epsilon_1$, and $s$ is the standard entropy density \cite{Kolb:1990vq}. 
The analogous asymmetry computed with $\epsilon^{vac}_1$ is denoted by $\eta^{vac}(M_1/T)$.

\begin{figure}[t]
	\begin{center}
		\includegraphics[width=\columnwidth]{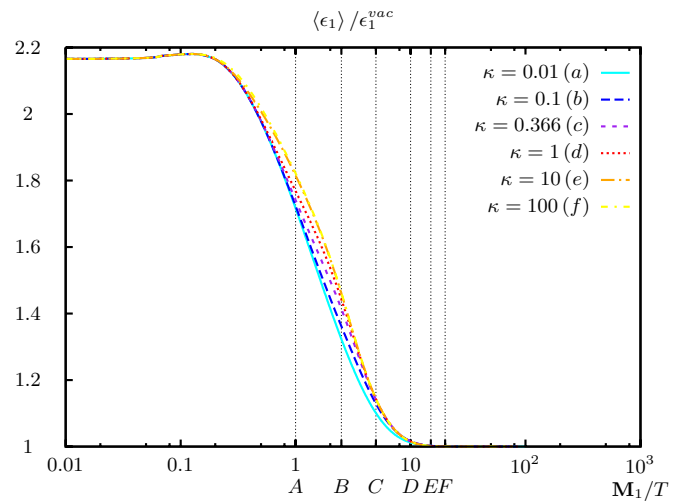}
	\end{center}
	\caption{\label{epsilonepsilonvacvsmt} 
		The ratio $\left<\epsilon_1\right>/\epsilon_1^{vac}$. The shape of the curves differs from that of 
		the corresponding graph in Fig.\,\ref{thermalEpsilonAveraged}, mainly because its computation 
		involves a finite chemical potential (which depends on $M_1/T$) here. Similar graphs can be 
		obtained by including a finite chemical potential in Eq.\,(\ref{epsilon_hierarchical_BE}).
	}
\end{figure} 
Figure \ref{epsilonepsilonvacvsmt} shows the numerical value of the ratio $\left<\epsilon_1\right>/\epsilon_1^{vac}$  
for various values of the washout parameter $\kappa\equiv \Gamma/H(M_1)=|g_1|^2m_{pl}/(4.5\cdot 16\pi\sqrt{g_*}M_1^3)$. 
The flattening for small $M_1/T$ as compared to the thermal equilibrium result in Fig.\,\ref{thermalEpsilonAveraged} 
is due to the finite chemical potential of $\bar{b}$. This shows that larger corrections could be obtained 
if additional interactions for $b$ and $\bar{b}$ are introduced which would allow one to start with smaller 
chemical potentials and hence lead to a stronger enhancement.

\begin{figure}[t]
	\begin{center}
		\includegraphics[width=\columnwidth]{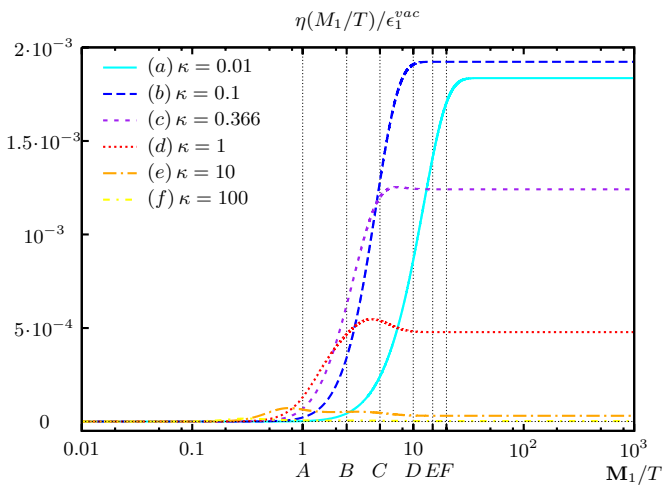}
	\end{center}
	\caption{\label{epsilonvsmt} 
		The asymmetry $\eta(M_1/T)$ with quantum corrections included. In the weak washout regime 
		(case $a$) the asymmetry is produced at smaller temperatures and it is not necessarily larger 
		than for larger washout factors (compare  $a$ and $b$).
	}
\end{figure} 
\begin{figure}[t]
	\begin{center}
		\includegraphics[width=\columnwidth]{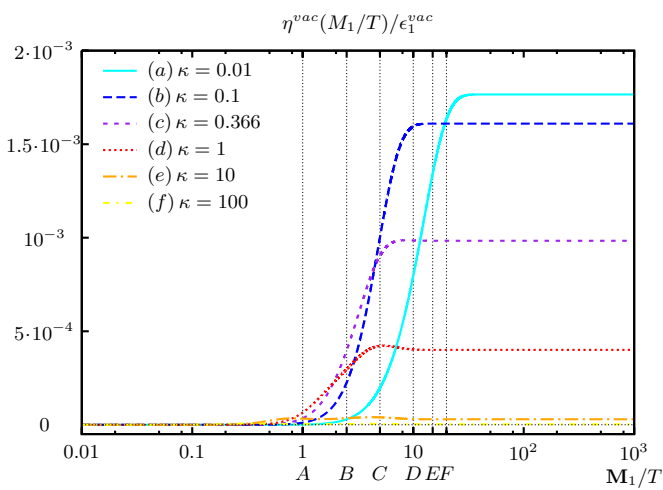}
	\end{center}
	\caption{\label{epsilonvsmtvac}  The asymmetry $\eta^{vac}(M_1/T)$ without quantum corrections.}
\end{figure} 
The buildup of the asymmetry with and without quantum corrections as a function 
of the inverse temperature is depicted in Figs.\,\ref{epsilonvsmt} and 
\ref{epsilonvsmtvac}. Comparing these figures one can verify the 
enhancement of the asymmetry at intermediate stages for larger washout factors 
(case $d$). Note also that due to the medium contribution to the \CP-violating 
parameter the generated asymmetry is not a monotonous  function of the washout parameter $\kappa$.

\begin{figure}[t]
	\begin{center}
		\includegraphics[width=\columnwidth]{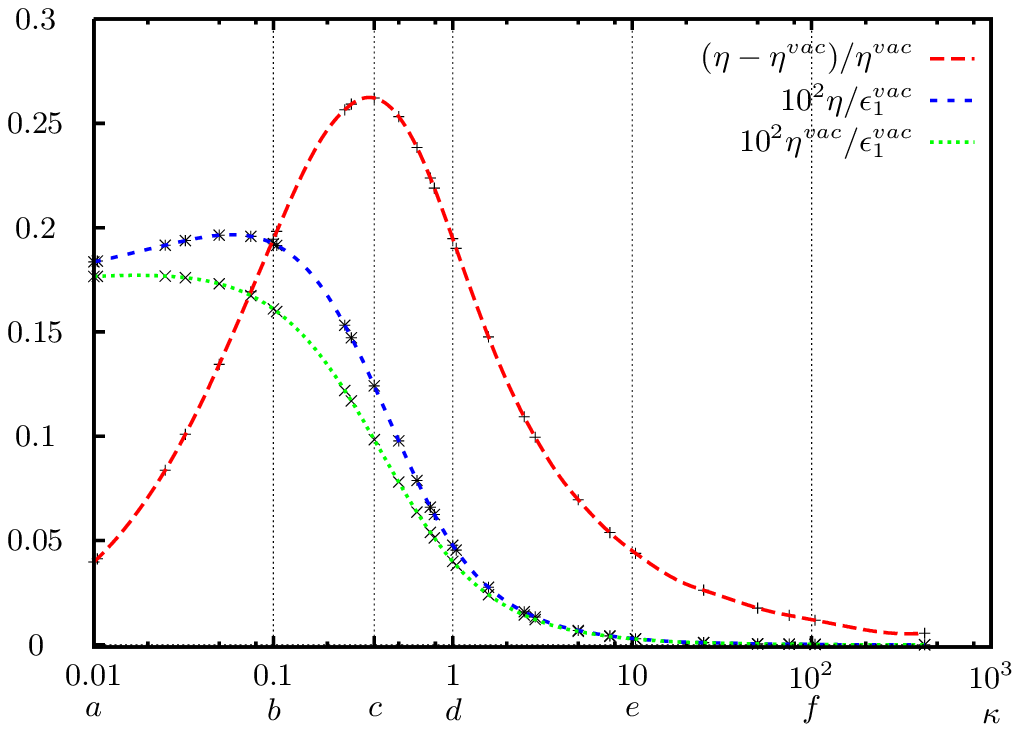}
	\end{center}
	\caption{\label{etavswashout} The final asymmetries and the ratio $(\eta-\eta^{vac})/\eta^{vac}$ over 
	washout factor $\kappa$. The cases $a$, $b$, $c$, $d$, $e$, $f$ correspond to washout factors 
	$\Ckappaa$, $\Ckappab$, $\Ckappac$, $\Ckappad$, $\Ckappae$, $\Ckappaf$. Case $c$ is close 
	to the maximum relative excess of the quantum-corrected results at $\kappa \simeq\Ckappacmaxreleta$. 
	In contrast to the usual results the final asymmetry does not take its maximum value for the 
	smallest washout factor. Instead, the asymmetry $\eta$ peaks at $\kappa\simeq\Ckappamaxeta$.}
\end{figure} 
The dependence of the resulting final asymmetries $\eta =\eta(M_1/T\rightarrow \infty)$ and
$\eta^{vac}=\eta^{vac}(M_1/T\rightarrow \infty)$ as well as  the ratio
$(\eta-\eta^{vac})/\eta^{vac}$  on the washout parameter is presented in   Fig.\,\ref{etavswashout}.
The asymmetry is always larger when quantum corrections are taken into account
compared to the results without  corrections (compare Sec.\,\ref{CPviol}).
The asymmetry $\eta$ has a maximum for moderate washout factors $\kappa\simeq \Ckappamaxeta$
in contrast to the usual result which has its maximum in the limit of zero
washout factor. Our interpretation of this result is as follows: For large
washout factors the enhancement of $\epsilon_1$ due to the quantum corrections
enhances the asymmetry generated by the decays only at intermediate stages,
because the same processes diminish the asymmetry in particular at late times
where the averaged asymmetry drops to smaller values (compare Fig.\,\ref{epsilonepsilonvacvsmt}). 
For small $\kappa$ the particles decay late, and the backreaction is largely suppressed 
so that the washout is ineffective. However the interval of integration in  
Eq.\,(\ref{epsilon hierarchical massless}) is located at relatively large momenta since the mass 
increasingly dominates $E_1=(M_1^2+\vec{p}^2)^\frac12$ as the momenta are
redshifted to smaller values. This means that the integration is over an
interval in which the distribution $\bar{f}$ becomes smaller and smaller.
This explains why the quantum corrections tend to zero for small $\kappa$.
For the same reasons the relative effect of the quantum corrections peaks
at a moderate $\kappa\simeq\Ckappacmaxreleta$ with about $26\%$. 

We note again that the size and the sign of the corrections depend on
the quantum statistics of the particles in the vertex loop and will
be different in a phenomenological scenario. Further plots and details
about the numerical algorithm can be found in Appendix\,\ref{app:NumericalDetails}.

\section{\label{Summary}Conclusions and outlook}

In this paper, we have studied leptogenesis in a simple toy model 
consisting of one complex and two real scalar fields in a top-down approach,  using the 
Schwinger--Keldysh/Kadanoff--Baym formalism as the starting point.
This treatment, based on nonequilibrium quantum field theory techniques, 
is motivated by the fact that it allows a unified description
of two key ingredients of leptogenesis, namely
deviation from thermal equilibrium and loop-induced \textit{CP} violation.
It has several important advantages in 
comparison to the canonical bottom-up (Boltzmann) approach. 
In particular, the full Kadanoff--Baym equations do not rely on the concept 
of quasiparticles and their collisions in the plasma. However, if the 
quasiparticle picture is applicable, as we have assumed here, the
Kadanoff--Baym formalism consistently accounts for the dependence of 
the quasiparticles' properties as well as scattering and decay rates on the state of the medium. 

The out-of-equilibrium dynamics of the quasiparticles 
is described by a system of approximate self-consistent kinetic 
equations -- \emph{quantum-corrected Boltzmann equations} -- which we have 
derived here starting from the full system of the Kadanoff--Baym equations.
We find that the structure of the quantum-corrected Boltzmann equations 
automatically ensures that no asymmetry is produced in thermal equilibrium. 
In other words there is no need for the real intermediate state subtraction, i.e.~the 
formalism is free of the double-counting  problem typical for the canonical approach.

One of the key quantities in leptogenesis is the \CP-violating parameter.
Earlier studies have shown that there are two sources of  \textit{CP} violation: 
self-energy and vertex contributions. In this work, we have concentrated on the latter one. 
We have found that for scalar fields the medium effects increase the 
vertex contribution to the \textit{CP}-violating parameter. At high 
temperatures it is up to an order of magnitude larger than in vacuum and asymptotically 
approaches the vacuum value as the temperature decreases.
This result can be traced back to a Bose enhancement of the
vertex loop correction. In the Maxwell--Boltzmann approximation, the
corresponding \textit{CP}-violating parameter is increased at most by a factor two. 
We would also like to note that, in the vacuum limit,  the \textit{CP}-violating 
parameter obtained via the Kadanoff--Baym formalism agrees with the value obtained
within the canonical formalism, as expected.

It is interesting that, contrary to the results obtained earlier in the 
framework of thermal field theory 
by replacing the zero-temperature propagators with finite temperature 
propagators in the matrix elements of the Boltzmann equation, 
the medium corrections 
depend only linearly on the particle number densities. Stated differently, only one of the 
internal lines in the vertex loop is ``thermal''. Moreover, the medium 
corrections to the vertex \textit{CP}-violating parameter depend only 
on the density of the toy baryons and are independent of the density 
of the ``Majorana'' particles. Since the decaying heavy particles as well as the 
intermediate on-shell states propagate with respect to the thermal bath's rest frame, 
the \textit{CP}-violating parameter in each individual decay depends on the momenta 
of the initial and final states.

We have  solved the system of the \textit{quantum-corrected
Boltzmann equations} numerically. Because of the medium corrections to the 
CP-violating parameter the asymmetry reaches its maximum value at a small
but finite value of the washout parameter $\kappa$, rather than
for $\kappa\rightarrow 0$, as it is the case in the canonical approach. 
To avoid the regime of Bose--Einstein condensation we have to assume that the species 
have rather large chemical potentials initially. 
This decreases the medium correction to the CP-violating parameter.
As a result the generated asymmetry differs from its value in the 
canonical formalism by approximately 26\%. However, in a scenario in which the chemical 
potentials are close to zero the deviation could reach the 100\% level. 
As has been mentioned above, our results differ from the results of the 
calculations performed in vacuum and in the framework of thermal field theory. 
Therefore, we argue that one should use the quantum-corrected Boltzmann equations 
(or the full Kadanoff--Baym equations). On the other hand, to obtain 
order of magnitude approximations, it seems safe to use the canonical 
approach in the present scenario.

The techniques described in this work can also be used to study quantum nonequilibrium
effects within phenomenological scenarios of leptogenesis. In particular, the
technical advantages of the Kadanoff--Baym formalism demonstrated above, like the absence of 
double-counting problems, are quite generic and therefore should not depend on the details of the model.
Furthermore, we expect that the difference in the size of the medium corrections, obtained from the
Kadanoff--Baym formalism and thermal field theory respectively, persists if our
approach is applied to such scenarios.

We would  like to stress again that, in this paper, we have considered 
only the vertex contribution to the \textit{CP}-violating parameter. The self-energy contribution 
is comparable to the vertex contribution or, 
in the case of resonant leptogenesis, considerably 
larger than the vertex contribution. We will study it in a forthcoming 
paper \cite{Garny:2009qn}. In addition in \cite{Garny:2009c} we plan to 
investigate the influence of a nonzero asymmetry on the CP-violating parameter.
Furthermore, for phenomenological models, thermal masses can become relevant. 
These can also consistently be described within the Kadanoff-Baym formalism.
Finally, we note that it would also be interesting to investigate
numerical solutions of the full set of Kadanoff--Baym equations without further
approximations as some of their properties cannot be included in Boltzmann-like equations for principle reasons. 
This is however beyond the scope of the present work.

\subsection*{Acknowledgements}
\noindent
This work was supported by the ``Sonderforschungsbereich'' TR27 and
by the ``cluster of excellence Origin and Structure of the Universe''.
We would like to thank J-S. Gagnon for useful discussions and
M.\,M.\,M\"uller for sharing his insights on nonequilibrium field 
theory.

\begin{appendix}

\section{\label{cpclassic}CP-violating parameter in the bottom-up approach}

In this appendix, we review the calculation of the vertex contribution to the
\textit{CP}-violating parameter in vacuum, $\epsilon_i^{\it vac} \equiv\left(\Gamma_{\psi_i \rightarrow{b}{b}}-\Gamma_{\psi_i \rightarrow\bar{b}\bar{b}}\right)/\left(\Gamma_{\psi_i \rightarrow{b}{b}}+\Gamma_{\psi_i \rightarrow\bar{b}\bar{b}}\right)$, 
in the conventional in-out formalism. It 
is generated by the interference of the tree-level and one-loop 
amplitudes (see Fig.\,\ref{interference}),
\begin{subequations}
	\label{amplitudesclassic}
	\begin{align}
		{\cal M}^{(0)}_{\psi_i\rightarrow bb}&=-ig_i^*\,,\nonumber\\
		{\cal M}^{(1)}_{\psi_i\rightarrow bb}&= ig_i g^*_j g^*_j\,
		{\textstyle\frac1{16\pi^2}}C_0(M_i^2,0,M_j^2)\,.\nonumber
	\end{align}
\end{subequations}
In the limit of massless toy baryons, the scalar 1-loop vertex 
three-point function $C_0$ is given by 
\cite{tHooft1979365,vanOldenborgh1990}:
\begin{align}
	\label{scalarvertexthreepointfunc_def}
	C_0(p_1^2,p_2^2,M_j^2)=\int\frac{({i\pi^2})^{-1}\,d^4q}{(q^2-M_j^2)(q+p_2)^2(q+p_2-p_1)^2}\,,
\end{align} 
where $p_1$ and $p_2$ are momenta of the decaying heavy 
particle and of one of the decay products, respectively. The tree-level and 
one-loop amplitudes of the decay process $\psi\rightarrow
\bar b \bar b$ differ from \eqref{amplitudesclassic}
only by  conjugation of the couplings. Therefore, 
at leading order, we obtain for the \CP-violating parameter: 
\begin{align}
	\label{epsilonvac_deriv1}
	\epsilon_i^{\it vac}=\frac{|{g_j}|^2}{8\pi^2}{\rm Im}\left(\frac{g_ig^*_j}{g^*_ig_j}\right){\rm Im}\, C_0(M_i^2,0,M_j^2)\,.
\end{align}
Substituting the result for the three-point 
function,
\begin{align}
	\label{scalarvertexthreepointfunc_res}
	C_0(M_i^2,0,M_j^2)=
	\frac{1}{M_i^2}\left[\mbox{Li}_2\left(1+\frac{M_i^2}{M_j^2}\right)-\frac{\pi^2}{6}\right]\,,
\end{align}
(with dilogarithm $\mbox{Li}_2$ as defined in \cite{vanOldenborgh1990})
into Eq.\,\eqref{epsilonvac_deriv1}, we obtain Eq.\,\eqref{epsilonvac}.
The same calculation can be performed within thermal quantum field theory
by using thermal propagators in Eq.\,(\ref{scalarvertexthreepointfunc_def}).
The result is \cite{Landsman:1986uw,Covi:1997dr}
\begin{align}
	\label{epsilon_th}
	\epsilon_i^{\it th}&(p_1,p_2)=-\frac1{8\pi}\frac{|g_j|^2}{M_i^2}{\rm Im}\left(\frac{g_ig^*_j}{g^*_ig_j}\right)\nonumber\\
	&\times\int\frac{d\Omega}{4\pi}\frac{1+\bar{f}^{\it th}_1+\bar{f}^{\it th}_2+2\bar{f}^{\it th}_1\bar{f}^{\it th}_2}{
	M_j^2/M_i^2+\frac12(1+\cos\theta)} + \dots \,,
\end{align}
where $\bar{f}^{\it th}_i = [\exp((E_{k_i}-\mu)/T)-1]^{-1}$, see Eq.\,(\ref{IntermedEnergy}).
The ellipsis denote similar contributions involving the distribution function of the
heavy scalar. These are suppressed in the hierarchical limit.

\section{\label{Complscal}Kadanoff--Baym formalism for the complex scalar field}

Here, we derive the Kadanoff--Baym and quantum-corrected Boltzmann equations 
for the complex scalar field.

\subsection{Schwinger--Dyson equation}

\begin{figure}[!ht]
	\begin{center}
		\includegraphics{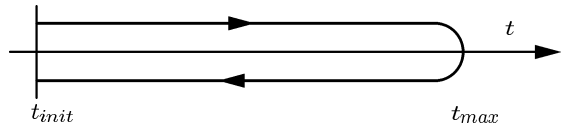}
	\end{center}
	\caption{\label{contour}Closed real-time path $\cal C$.}
\end{figure}

Our starting point is the generating functional for Green's 
functions~\cite{PhysRevD.10.2428}:
\begin{align}
	\label{genfunct_complex}
	{\cal Z}[J,K]=
	{\textstyle\int}\mathscr{D}b\mathscr{D}\bar{b}
	\exp[i(S+J\bar{b}+\bar{J}b+\bar{b}Kb)]\,,
\end{align}
where the field and the external sources are defined on the
the positive and negative branches of the Schwinger-Keldysh closed real-time contour shown in Fig.\,\ref{contour} \cite{Schwinger:1960qe,Keldysh:1964ud,Bakshi:1963ab,Danielewicz:1982kk,Chou:1984es,Calzetta:1986cq}.
The scalar products of the local and bilocal sources $J(x)$ 
and $K(x,y)$ and the field are defined as invariant configuration space integrals 
\cite{Calzetta:1986ey,Hohenegger:2008zk}.
Furthermore, we use the compact notation of Ref.~\cite{Danielewicz:1982kk}
for contour integrals over the closed real-time path.
Note that the sources are now complex 
functions. The requirement that the last term in \eqref{genfunct_complex} 
be real implies that $K(x,y)=K^*(y,x)$.

The functional derivatives of the generating functional for connected Green's functions,
\begin{align}
	{\cal W}[J,K]=-i\ln {\cal Z}[J,K]\,,
\end{align}
with respect to the external sources read
\begin{subequations}
	\label{WDerivatives}
	\begin{align}
		\label{JbarDerivative}
		\frac{\partial {\cal W}[J,K]}{\partial J(x)}&=\bar{B}(x)\,,\\
		\frac{\partial {\cal W}[J,K]}{\partial K(x,y)}&=
		{\textstyle\frac12}[D(y,x)+\bar{B}(x)B(y)]\,.
	\end{align}
\end{subequations}
$B$ and $D$ denote the expectation value and the  propagator of the field respectively. 
The derivative of $\cal W$ with respect to $\bar J$ is just the complex conjugate of \eqref{JbarDerivative}.

Performing a Legendre transform of the generating functional for connected Green's functions, 
we obtain the effective action
\begin{align}
	\hskip -1.5mm
	\Gamma[D,B]\equiv {\cal W}[J,K]
	-J\bar{B}-\bar{J}B-\tr[KD]-\bar{B} K B .
\end{align}
Making use of the chain rule and Eqs. \eqref{WDerivatives}, we find for functional 
derivatives of the effective action
\begin{subequations}
	\label{DiffGamma_complex}
	\begin{align}
		\frac{\delta \Gamma[D,B]}{\delta \bar{B}(x)}&=-J(x)
		-{\textstyle\int}\dg z~ K(x,z)B(z)\,,\\
		\label{DiffGammaD}
		\frac{\delta \Gamma[D,B]}{\delta D(x,y)}&=
		-K(y,x)\,.
	\end{align}
\end{subequations}

Next, we shift the complex field by its expectation value $b\rightarrow b+B$.
Since the integration measure in the path integral is translationally
invariant, the effective action can be rewritten in the form
\begin{align}
	\Gamma[D,B]=&-i\ln {\textstyle\int} \mathscr{D}b \mathscr{D}\bar{b}
	\exp [i(S +\bar{J}b+J\bar{b}+\bar{b} Kb)]
	\nonumber\\
	&+S_{cl}[B]-\tr[KD]\,.
\end{align}
Now, we tentatively write the effective action in the form~\cite{PhysRevD.10.2428}
\begin{align}
	\label{D2deff}
	\Gamma[D,B]&\equiv S_{cl}[B]+i
	\ln\det\left[D^{-1}\right]+i\,\tr\left[\mathscr{D}^{-1}D\right]\nonumber\\
	&+\Gamma_2[D,B]\,,
\end{align}
thus defining the functional $\Gamma_2$. 

The third term on the right-hand side of \eqref{D2deff}
is given by a convolution of the field propagator $D$ and the 
free inverse propagator $\mathscr{D}^{-1}$. Its differentiation with respect 
to $D(y,x)$ gives \cite{Hohenegger:2008zk}
\begin{align}
	\mathscr{D}^{-1}(x,y)=i(\square_x+m^2)\,\delta^g(x,y)\,.
\end{align}
The functional derivative of the second term on the 
right-hand side of \eqref{D2deff} can be obtained 
upon use of 
\begin{align}
	\label{DinvProp}
	\textint \dg z\,D^{-1}(x,z)D(z,y)=\delta^g(x,y)\,,
\end{align}
and is given by $-iD^{-1}(y,x)\,$. Consequently, we obtain
\begin{align}
	\label{D2diff}
	\frac{\delta \Gamma[D,B]}{\delta D(x,y)}=
	&-iD^{-1}(y,x)+i\,\mathscr{D}^{-1}(y,x)
	+\frac{\delta \Gamma_2[D,B]}{\delta D(x,y)}\nonumber\\
	=&-K(y,x)\,.
\end{align}
Physical reality corresponds to vanishing sources.\footnote{To be
precise, within nonequilibrium field theory, this is only true for times $x^0,y^0>t_{\it init}$. The local and
bi-local sources supported at $x^0=y^0=t_{\it init}$ formally
encode the information about the (Gaussian) initial state (see e.g.~\cite{Berges:2004yj}).
However, these sources do not appear explicitly in the Kadanoff-Baym equations, and therefore
we omit them here.} Therefore, Eq. \eqref{D2diff} can be rewritten in the form
\begin{align}
	\label{SDeqsC}
	D^{-1}(x,y)=\mathscr{D}^{-1}(x,y)-\Sigma(x,y)\,,
\end{align}
where the self-energy is defined by
\begin{align}
	\label{selfenergydeffC}
	\Sigma(x,y)\equiv i\frac{\delta \Gamma_2[D,B]}{\delta D(y,x)}\,.
\end{align}
Note that the factor two in the definition of the self-energy 
\cite{Lindner:2005kv,Hohenegger:2008zk} is absent, just as one would expect 
for a complex field.

\subsection{Kadanoff--Baym equations}

Convolving the Schwinger--Dyson equation \eqref{SDeqsC} with $D$ 
from the right and using Eq.\,\,\eqref{DinvProp}, we obtain
\begin{align}
	\label{convolutionC}
	i[\square_x+m^2]D(x,y)&=\delta_g(x,y)\nonumber\\
	&+\textint\dg z\,\Sigma(x,z)D(z,y)\,.
\end{align}
Next, following the usual procedure, we represent the time-ordered 
propagator as a linear combination of the statistical propagator 
and spectral function:
\begin{align}
	\label{DfDr}
	D(x,y)&=D_F(x,y)-{\textstyle\frac{i}{2}}\sign_{\cal C}(x^0-y^0)D_\rho(x,y)\,,
\end{align}
where $\sign_{\cal C}$ denotes the signum function with respect to time ordering
along the closed time path, and
\begin{subequations} 
	\label{DrhoDFdeff}
	\begin{align}
		\label{Df}
		D_F(x,y)&\equiv{\textstyle\frac12}\langle [b(x),\bar{b}(y)]_{+}\rangle\,,\\
		\label{Drho}
		D_\rho(x,y)&\equiv i\langle [b(x),\bar{b}(y)]_{-}\rangle\,,
	\end{align}
\end{subequations}
where the subscripts ``$+$'' and ``$-$'' denote the anticommutator and the 
commutator of the fields.

To find out how $D_F$ and $D_\rho$ behave under complex 
conjugation let us introduce 
\begin{subequations}
	\label{gtrlesscomponents}
	\begin{align}
		D_>(x,y)&\equiv\langle b(x)\bar{b}(y)\rangle=
		\Tr[\mathscr{P}\,b(x)\bar{b}(y)]\,,
		\\
		D_<(x,y)&\equiv\langle \bar{b}(y) b(x)\rangle=
		\Tr[\mathscr{P}\,\bar{b}(y)b(x)]\,.
	\end{align}
\end{subequations}
Using the Hermiticity of the density matrix $\mathscr{P}$
and the cyclic invariance of the trace, we obtain
\begin{align}
	\label{gtrless_properties}
	D^*_>(x,y)=D_>(y,x),\quad D^*_<(x,y)=D_<(y,x)\,.
\end{align}
Consequently 
\begin{align}
	\label{DFDrhoProp}
	D^*_F(x,y)=D_F(y,x),\quad D^*_\rho(x,y)=-D_\rho(y,x)\,.
\end{align}
Analogous relations also hold for the spectral 
and statistical components of the self-energy.

The local part of the self-energy is proportional to the Dirac $\delta$ function  
and can be absorbed in the effective mass of the field,
$m^2(x)\equiv m^2+\Sigma^{\it loc}(x,x)\,,$
whereas the remaining part of the self-energy can be split 
into a spectral part $\Sigma_\rho$ and a statistical part $\Sigma_F$
in a complete analogy to \eqref{DfDr}.

Because of the $\sign$ function, the action of the Laplace--Beltrami 
operator on \eqref{DfDr} gives rise
to a product of $g^{00}\delta(x^0,y^0)$ and $\nabla^x_0D_\rho(x,y)$.
Upon use of the definition of the spectral function and 
canonical commutation relations for a complex scalar field
this product reduces to the generalized $\delta$ function 
$\delta^g(x,y)$ \cite{Hohenegger:2008zk}, which  cancels 
 the $\delta$ function on the right-hand side of 
\eqref{convolutionC}.

Separating spectral and statistical components in Eq.\,(\ref{convolutionC}),
we obtain a system of  Kadanoff--Baym equations very 
similar to that for the real scalar field \cite{Hohenegger:2008zk}:
\begin{align}
	\hspace{-2mm}
	\label{DFeq}
	[\square_x+m^2(x)]D_F(x,y)&=
	{\textstyle\int\limits^{y^0}_0}\dg z\,\Sigma_F(x,z)D_\rho(z,y)\nonumber\\
	&-{\textstyle\int\limits^{x^0}_0}\dg z\,\Sigma_\rho(x,z)D_F(z,y),\\
	\label{Drhoeq}
	[\square_x+m^2(x)]D_\rho(x,y)&=
	{\textstyle\int\limits_{x^0}^{y^0}}\dg z\,\Sigma_\rho(x,z)D_\rho(z,y)\,.
\end{align}
One should, however, keep in mind that the functions in \eqref{DFeq}
and \eqref{Drhoeq} are \textit{complex}. That is, we get four equations 
for the real and imaginary components of the spectral function and the 
statistical propagator.
The complete information about the (Gaussian) initial state specified
at the initial time $t_{\it init}\equiv 0$ enters via the initial conditions of the
two-point functions $D_F$, $\partial_{x^0}D_F$, $\partial_{y^0}D_F$ and
$\partial_{x^0}\partial_{y^0}D_F$ evaluated at $x^0=y^0=t_{\it init}$.
The corresponding initial conditions for the spectral function are
fixed by the equal-time commutation relation of the complex field,
as for the real case~\cite{Berges:2004yj}.

We note that the proper renormalization of the Kadanoff-Baym
equations (\ref{DFeq},\ref{Drhoeq}) would require one to also take 
non-Gaussian correlations of the initial state into
account~\cite{Garny:2009ni,Borsanyi:2008ar}. However, the
derivation of quantum-corrected Boltzmann equations considered
here involves the limit $t_{\it init}\rightarrow -\infty$, and
therefore the initial correlations should have a negligible effect.

\subsection{Quantum kinetics}
The Kadanoff--Baym equation for the statistical propagator (spectral
function) can be rewritten in terms of the advanced and retarded 
propagators, $D_R$ and $D_A$:
\begin{align}
\label{KBeqsQKC}
[&\square_x+m^2(x)]D_{F(\rho)}(x,y)=-\textint
\dg z
\theta(z^0)\nonumber\\
&\times[\Sigma_{F(\rho)}(x,z)D_A(z,y)+\Sigma_R(x,z)D_{F(\rho)}(z,y)]\,.
\end{align}
Because of \eqref{DFDrhoProp}, the retarded and advanced 
propagators are related by 
\begin{align}
\label{DRDAconj}
D_R(x,y)&\equiv \theta(x^0-y^0)D_\rho(x,y)\nonumber\\
&=-\theta(x^0-y^0)D^*_\rho(y,x)=D^*_A(y,x)\,.
\end{align}
Therefore, after interchange of $x$ and $y$ in \eqref{KBeqsQKC} and 
complex conjugation of the resulting equation, we find
\begin{align}
\label{DFQKichng}
[&\square_y+m^2(y)]D_{F(\rho)}(x,y)=-\textint
\dg z
\theta(z^0)\nonumber\\
&\times[D_R(x,z)\Sigma_{F(\rho)}(z,y)+D_{F(\rho)}(x,z)\Sigma_A(z,y)]\,.
\end{align}
Since \eqref{DFQKichng} has been obtained from \eqref{KBeqsQKC} 
by reversible transformations, a solution of \eqref{KBeqsQKC}
is also a solution of \eqref{DFQKichng}. Consequently, it is also
a solution of the sum (which will be referred to as \textit{constraint}
equation) and the difference (which will be referred to as \textit{kinetic} 
equation) of \eqref{KBeqsQKC} and \eqref{DFQKichng}.

To analyze the constraint and kinetic equations it is convenient 
to introduce the center and relative coordinates, $X$ and $s$ 
\cite{PhysRevD.32.1871}. In terms of the center and relative coordinates 
relations \eqref{DFDrhoProp} can be rewritten in the form 
\begin{align*}
{D}^*_F(X,s)={{D}_F(X,-s)},\quad 
{D}^*_\rho(X,s)=-{{D}_\rho(X,-s)}\,.
\end{align*}
Consequently, even in the case of a complex scalar field,  
the Wigner transforms of the spectral function and statistical 
propagator,
\begin{subequations}
\label{WignerTrafos}
\begin{align}
\label{DFXp}
{D}_{F}(X,p)&=\sqrt{-g}_X\textint d^4s\, e^{ips}{{D}_F(X,s)}\,,\\
\label{DrhoXp}
{D}_{\rho}(X,p)&=-i\sqrt{-g}_X\textint d^4s\, e^{ips}{{D}_\rho(X,s)}\,,
\end{align}
\end{subequations}
are real-valued functions. The Wigner transforms of the 
retarded and advanced propagators are defined analogously 
to \eqref{DFXp}. From \eqref{DRDAconj} it then follows that 
relation
\begin{align}
\label{DADRrel}
{D}_{A}(X,p)&={D}^*_R(X,p)
\end{align}
also holds for a complex scalar field. Another very useful 
relation,  
\begin{align}
\label{DAminDR}
{D}_{R}(X,p)&-{D}_{A}(X,p)=i{D}_{\rho}(X,p)\,,
\end{align} 
results from the definitions \eqref{DRDAconj}  and 
\eqref{WignerTrafos} and the equality $\theta(s^0)+\theta(-s^0)=1$.
Equations \eqref{DADRrel} and \eqref{DAminDR} in particular imply  that 
\begin{align}
{D}_{R(A)}(X,p)={D}_{h}(X,p)\pm
 {\textstyle\frac{i}{2}} {D}_{\rho}(X,p),\quad 
\end{align}
where ${D}_{h}(X,p)\equiv \Re\, {D}_{R}(X,p)$ has been 
introduced. A similar relation also holds for the retarded and 
advanced self-energies.

Let us now subtract \eqref{DFQKichng} from  \eqref{KBeqsQKC} and then
Wigner transform the left- and right-hand sides of the resulting
equation. Furthermore, we send the initial time to the infinite
past, $t_{\it init}\rightarrow -\infty$, i.e.~we drop the functions $\theta(z^0)$
on the right-hand sides of Eqs.~\eqref{KBeqsQKC} and~\eqref{DFQKichng}. Physically,
this means that we neglect the effects of initial correlations. Additionally, we
perform a gradient expansion with respect to the central coordinate $X$.
Proceeding as in \cite{Hohenegger:2008zk}, we obtain a kinetic 
equation for the spectral function. To linear order in the gradients 
it reads
\begin{align}
\label{DrhoQKequation}
\hspace{-2.8mm}
\{{\omega}(X,p),{D}_{\rho}(X,p)\}_{PB}
=\{{\Sigma}_{\rho}(X,p),{D}_{h}(X,p)\}_{PB},
\end{align}
where we have introduced 
\begin{align}
\label{omegadef}
{\omega}(X,p)\equiv g^{\mu\nu}p_\mu p_\nu-m^2(X)-{\Sigma}_{h}(X,p)\,,
\end{align}
and the Poisson brackets are defined by \cite{Hohenegger:2008zk}
\begin{align}
\label{PoissonBrackets}
\{{A}(X,p),{B}(X,p)&\}_{PB}\equiv
\frac{\partial}{\partial p_\alpha} {A}(X,p) {\cal D}_\alpha{B}(X,p)\nonumber\\
&-{\cal D}_\alpha {A}(X,p) \frac{\partial}{\partial p_\alpha}{B}(X,p)\,.
\end{align}

Wigner transforming the sum of \eqref{DFQKichng} and  \eqref{KBeqsQKC}, 
we obtain the constraint equation for the spectral function. To linear order in the gradients it is an \textit{algebraic} equation: 
\begin{align}
\label{Drho_spectr}
{\omega}(X,p)  D_{\rho}(X,p)=\Sigma_{\rho}(X,p) D_h(X,p)\,.
\end{align}

To close the system and to analyze the spectrum, we also need the 
equations for the retarded and advanced propagators. They can 
be obtained from \eqref{KBeqsQKC} and \eqref{DFQKichng} upon 
use of the definitions of $D_R$ and $D_A$ and the canonical 
commutation relations:
\begin{align}
\label{DRAQK}
[\square_x+m^2(x)]&D_{R(A)}(x,y)=\delta^g(x,y)\nonumber\\
&-\textint \dg z\, \Sigma_{R(A)}(x,z)D_{R(A)}(z,y)\,,\\
\label{DRAQK2}
[\square_y+m^2(y)]&D_{A(R)}(x,y)=\delta^g(x,y)\nonumber\\
&-\textint \dg z\, D_{A(R)}(x,z)\Sigma_{A(R)}(z,y)\,.
\end{align}
Wigner  transforming the \textit{difference} of \eqref{DRAQK} and \eqref{DRAQK2} 
and subtracting \eqref{DrhoQKequation}, we obtain the kinetic equation 
for real part of the retarded and advanced propagators:
\begin{align}
\label{DhQKequation}
\{{\omega}(X,p),{D}_{h}(X,p)&\}_{PB}=\nonumber\\
&-{\textstyle\frac14}\{{\Sigma}_{\rho}(X,p),{D}_{\rho}(X,p)\}_{PB}\,.
\end{align}
Wigner transforming the \textit{sum} of \eqref{DRAQK} and \eqref{DRAQK2}  
and subtracting \eqref{Drho_spectr}, we obtain the second constraint equation:
\begin{align}
\label{DR_spectr}
{\omega}(X,p){D}_{h}(X,p)=-1-
{\textstyle\frac14}{\Sigma}_{\rho}(X,p){D}_{\rho}(X,p)\,.
\end{align} 
The solution of the system of  constraint equations 
\eqref{Drho_spectr} and \eqref{DR_spectr} reads
\begin{subequations}
\label{Dsolutions}
\begin{align}
\label{Drhosolution}
{D}_{\rho}(X,p)&=\frac{-{\Sigma}_{\rho}(X,p)}{\omega^2(X,p)+
{\textstyle\frac14}{\Sigma}^2_\rho(X,p)}\,,
\\
{D}_{h}(X,p)&=\frac{{\omega}(X,p)}{{\Sigma}_{\rho}(X,p)}
{D}_{\rho}(X,p)\,.
\end{align}
\end{subequations}
As can be checked by substitution, solution \eqref{Dsolutions}
is also solution of the \textit{kinetic} equations \eqref{DrhoQKequation}
and \eqref{DhQKequation}. In other words, to linear order in the 
gradients we have analytic expressions for the spectral function and 
retarded (advanced) propagators. The spectral function has a sharp 
peak on the mass shell, i.e.~for ${\omega}(X,p)=0$. 
The height and exact shape of the peak are time-dependent.

Proceeding in a similar way, we can derive the kinetic 
\begin{align}
\label{DFQKequation}
\{\omega&(X,p),{D}_{F}(X,p)\}_{PB}
=\{{\Sigma}_{F}(X,p),{D}_{h}(X,p)\}_{PB}
\nonumber\\
&+{D}_{F}(X,p){\Sigma}_{\rho}(X,p)-
{\Sigma}_{F}(X,p){D}_{\rho}(X,p)\,,
\end{align}
and the constraint,
\begin{align}
\label{DFconstraint}
&{\omega}(X,p){D}_{F}(X,p)={\textstyle\frac14}\{{\Sigma}_{F}(X,p),
{D}_{\rho}(X,p)\}_{PB}\\
&+{\textstyle\frac14}\{{D}_{F}(X,p),{\Sigma}_{\rho}(X,p)\}_{PB}
+{\Sigma}_{F}(X,p){D}_{h}(X,p)\nonumber
\end{align}
equations for the statistical propagator. The constraint equation
is no longer algebraic and cannot be, generally speaking, solved analytically.
However, if the system is in thermal equilibrium, then 
all the quantities are constant in time and space and 
the Poisson brackets in \eqref{DFconstraint} vanish identically. 
The solution of the resulting equation reads
\begin{align}
\label{DFequilibrium}
{D}^{eq}_{F}(p)=
\frac{{\Sigma}_{F}(p)}{{\Sigma}_{\rho}(p)} {D}^{eq}_{\rho}(p)\,.
\end{align}
That is, we have obtained the fluctuation-dissipation relation 
from the constraint equation \eqref{DFconstraint}. As can be checked 
by substitution, in equilibrium \eqref{DFequilibrium} is indeed a 
solution of \eqref{DFQKequation}. Furthermore, using 
\eqref{DfDr} and the Kubo-Martin-Schwinger periodicity condition we find 
\cite{Berges:2004yj} 
\begin{align}
\label{fluct_diss}
{D}^{eq}_F(p)=\left[\vphantom{{\textstyle\frac12}}\hspace{-1mm}\right.
f^{eq}(p)+\left.{\textstyle\frac12}\right]
{D}^{eq}_\rho(p)\,,
\end{align}
where $f^{(eq)}$ is the Bose--Einstein distribution function.

\subsection{Quantum-corrected Boltzmann equations}

The spectral function for the complex scalar field \eqref{Drhosolution} has a 
Breit--Wigner shape and peaks at $\omega=0$. The height of the 
peak is inversely proportional to the spectral self-energy and tends 
to infinity in the limit of vanishing coupling constant. Since, furthermore, 
the area under ${D}_{\rho}(X,p)$ is constant \cite{Hohenegger:2008zk} to a first
approximation it can be replaced by a Dirac $\delta$ function:
\begin{align}
\label{QPapprox}
{D}_{\rho}(X,p)=2\pi\,\sign(p_0) \, 
\delta\left(g_{\mu\nu} p^\mu p^\nu-m^2\right)\,.
\end{align}
Equation \eqref{QPapprox} is referred to as \textit{quasiparticle  
approximation}. 

Strictly speaking, the quasiparticle approximation is sufficient 
only for the analysis of lowest-order processes. In the model under 
consideration this includes the tree-level decay and tree-level 
scattering processes, which are obtained at the order ${\cal O}(g^2$) and 
${\cal O}(\lambda^2)$, respectively (see Fig.\,\ref{processes}).
To ensure the consistency of the description beyond  leading order one 
should use the so-called \textit{extended quasiparticle 
approximation} \cite{PhysRevB.52.14615,PhysRevC.46.1687,PhysRevC.48.1034,
PhysRevC.64.024613,CondMatPhys2006_9_473,JPhys2006_35_110}. The 
extended quasiparticle approximation for the complex scalar field 
would allow us, for instance, to describe $\psi b\rightarrow b\rightarrow 
\psi b$ scattering processes, which are of order  ${\cal O}(g^4)$.
In this paper we are primarily interested in the three-loop vertex 
diagram, whose contribution to the effective action contains the fourth
power of the coupling $g$. Using the quasiparticle approximation for the 
propagators in the vertex diagram thus induces contributions which are already
of order  ${\cal O}(g^4)$. Consequently, the extended quasiparticle
approximation would additionally induce contributions of higher order
in the coupling constants. Therefore, the calculation of the leading 
order contribution to the vertex \textit{CP}-violating parameter does not 
require us to go beyond the quasiparticle approximation. 

As has been argued in \cite{Hohenegger:2008zk}, in the same approximation
one can also neglect the Poisson brackets in the kinetic equations, which
physically corresponds to the \textit{Stosszahlansatz} of Boltzmann. This 
leads to a simple kinetic equation for the spectral function:
\begin{align}
\label{BEGrho}
p^\alpha {\cal D}_\alpha {D}_{\rho}(X,p)&=0\,.
\end{align}
Let us note, that the quasiparticle approximation for the spectral function
\eqref{QPapprox} is consistent with \eqref{BEGrho}.

Neglecting the Poisson brackets on the right-hand side of 
\eqref{DFQKequation}, we obtain the Boltzmann equation for 
the statistical propagator:
\begin{align}
\label{DBlzmnPrelim}
p^\alpha{\cal D}_\alpha {D}_{F}(X,p)
={\textstyle\frac12}[
{D}_{F}&(X,p){\Sigma}_{\rho}(X,p)\nonumber\\
&-{\Sigma}_{F}(X,p){D}_{\rho}(X,p)]\,.
\end{align}
Motivated by the fluctuation-dissipation relation \eqref{fluct_diss}, 
we trade the statistical propagator for the one-particle number density:
\begin{align}
\label{KBAnsatz_complex}
{D}_{F}(X,p)=\left[f(X,p)+{\textstyle\frac12}\right]
{D}_{\rho}(X,p)\,.
\end{align}
In view of \eqref{BEGrho}, we can then rewrite \eqref{DBlzmnPrelim} as 
an equation for the phase-space  distribution function $f(X,p)$:
\begin{align}
\label{DBlzmn}
[\,p^\alpha {\cal D}_\alpha &f(X,p)]{D}_{\rho}(X,p)\\
&={\textstyle\frac12}[{\Sigma}_{>}(X,p){D}_{<}(X,p)
-{D}_{>}(X,p){\Sigma}_{<}(X,p)]\,,\nonumber
\end{align}
where we have introduced
\begin{align}
{D}_{\gtrless}(X,p)={D}_{F}(X,p)\pm\textstyle\frac12{D}_{\rho}(X,p)\,.
\end{align}
Equation (\ref{DBlzmn}) is very similar to the Boltzmann equation for a real 
scalar field \cite{Hohenegger:2008zk}. There is, however, an important 
difference. For negative values of $p_0$ the distribution function $f$ 
describes \textit{antiparticles}:
\begin{align}
\label{partantipartrel}
f(X,-p)\equiv-[\bar{f}(X,p)+1]\,.
\end{align}
In other words, Eq.\,\eqref{DBlzmn} describes the 
time evolution of both particles and antiparticles. One can obtain an explicit
equation for $\bar{f}$ by changing the sign of $p_0$:
\begin{align}
\label{DBlzmn_antipart}
[\,p^\alpha {\cal D}_\alpha &\bar{f}(X,p)]{D}_{\rho}(X,p)\\
&={\textstyle\frac12}[{\bar\Sigma}_{>}(X,p){\bar D}_{<}(X,p)
-{\bar D}_{>}(X,p){\bar \Sigma}_{<}(X,p)]\,,\nonumber
\end{align}
where we have introduced ${\bar\Sigma}_{\gtrless}(X,p)\equiv 
{\Sigma}_{\lessgtr}(X,-p)$ and taken into account that in the 
quasiparticle approximation ${D}_{\rho}(X,-p)=-{D}_{\rho}(X,p)$.

\section{\label{selfenergies}Calculation of the self-energies}
The 2PI effective action is given by the sum of all 2PI diagrams with 
vertices as given by the interaction Lagrangian and internal lines 
representing the complete connected propagators \cite{Berges:2004yj}. 
The structure of the terms of the effective action can be read off the 
diagrams in Fig.\,\ref{diagrams}:
\begin{subequations}
\begin{align}
i\Gamma^{(a)}_2=&{\textstyle -\frac{i}{2}} \lambda \intg{x}
D^2(x,x)\,,\\
i\Gamma^{(b)}_2=&{\textstyle -\frac{1}{8}}\lambda^2\intg{xy}
 D^2(x,y)D^2(y,x)\,,\\
i\Gamma^{(c)}_2=&-{\textstyle\frac14}\, g_m g_n^* \intg{xy}G^{mn}(x,y) D^2(x,y)\nonumber\\
&-{\textstyle\frac14}\, g^*_m g_n \intg{xy}G^{mn}(x,y) D^2(y,x)\,,\\
i\Gamma^{(d)}_2=& {\textstyle\frac14} g_i g_j
g^*_m g^*_n \intg{xyvu}  G^{ij}(x,y) G^{mn}(v,u)\nonumber\\
 & \hspace{6mm} \times D(y,v)D(x,v)
 D(y,u)D(x,u)\,,
\end{align}
\end{subequations}
where, to shorten the notation, we have introduced 
\begin{align*}
\intg{x_1\ldots\,x_n}\equiv {\textstyle\int} \dg{x_1}\ldots\,\dg{x_n}\,.
\end{align*}
The self-energies of the complex scalar field are obtained by functional 
differentiation of the effective action with respect to the two-point 
correlation function: 
\begin{align}
\Sigma(x,y)\equiv i\frac{\delta \Gamma_2[D,G]}{\delta D(y,x)}\,.
\end{align}
Differentiating the individual contributions to the effective action, 
we obtain
\begin{subequations}
\label{Sigma}
\begin{align}
\Sigma^{(a)}(x,y)&=-i\delta^g(x,y)\lambda D(x,x)\,,\\
\Sigma^{(b)}(x,y)&=-{\textstyle\frac12}\lambda^2 D^2(x,y)D(y,x)\,,\\
\Sigma^{(c)}(x,y)&=-g_i g_j^*G^{ij}(y,x)D(y,x)\,,\\
\label{Sigma_d}
\Sigma^{(d)}(x,y)&=g_i g_j g_m^* g_n^* \intg{vu} G^{mn}(x,v) G^{ij}(y,u)\nonumber\\
&\hspace{16mm}\times  D(y,v)D(u,v)D(u,x)\,.
\end{align}
\end{subequations}
The components of the self-energy of the system of real scalar fields 
are obtained upon functional differentiation of the effective action 
with respect to the  components of the correlation function:
\begin{align}
\Pi^{ij}(x,y)\equiv 2i\frac{\delta \Gamma_2[D,G]}{\delta G^{ji}(y,x)}\,.
\end{align}
The result of the differentiation reads
\begin{subequations}
\label{Pi}
\begin{align}
\Pi^{(c)}_{ij}(x,y)&=-{\textstyle\frac12}g_i g^*_j D^2(x,y)
-{\textstyle\frac12}g^*_i g_j D^2(y,x)\,,\\
\Pi^{(d)}_{ij}(x,y)&={\textstyle\frac12}\intg{vu}\,
G^{mn}(v,u) \nonumber\\
\times [\,g_i & g_j  g_m^* g_n^* D(x,v)D(x,u)D(y,v)D(y,u) \nonumber\\
+g^*_i& g^*_j  g_m g_n  D(v,x)D(u,x)D(v,y)D(u,y)]\,.
\end{align}
\end{subequations}
The next step is to derive the spectral and statistical components of
the self-energies \eqref{Sigma} and \eqref{Pi}. Upon use of the 
decomposition \eqref{DfDr} and of the analogous decomposition of
the propagators of the real scalar field, one easily
obtains a corresponding decomposition of the self-energies $\Sigma^{(b)}$, $\Sigma^{(c)}$ 
and $\Pi^{(c)}$  into the statistical and spectral 
components. Linear combinations of the resulting expressions 
are presented in Eqs.\,\eqref{Sigmab}, \eqref{Sigmac}, and \eqref{Pic}. 
The calculation of the spectral and statistical components of 
$\Sigma^{(d)}$ and $\Pi^{(d)}$, which contain two integrations over
space-time, is more involved (see also \cite{Carrington:2004tm}).
Decomposing the two-point correlation
functions in Eq.\,\eqref{Sigma_d} into the statistical and spectral 
components, we get 32 terms. Each of the terms must be integrated over 
the closed time path $\cal C$, see Fig.\,\ref{contour}.
It is helpful to use relations like the following,
\begin{align}
	\lefteqn{ \int_{\mathcal{C}} du^0 \, \sign_\mathcal{C}(x^0-u^0) \, \sign_\mathcal{C}(u^0-y^0) \, D_\rho(x,u)D_\rho(u,y)} \nonumber\\
	& =  2 \,\sign_\mathcal{C}(x^0-y^0) \, \int_{y^0}^{x^0} du^0 \, D_\rho(x,u)D_\rho(u,y)\,.
\end{align}
One then finds that ten terms vanish
upon integration over the contour: one term which does not contain any
$\sign_{\cal C}$ functions; five terms which contain only one $\sign_{\cal C}$ 
function; two terms which contain a product of two $\sign_{\cal C}$ 
functions both depending only on one of the integration variables, 
$u$ or $v$; and finally two terms which contain a product of three 
$\sign_{\cal C}$ functions but depend only on one of the ``external''
arguments, $x$ or $y$. 
\begin{figure}[!ht]
 \begin{center}
\includegraphics[width=0.30\textwidth]{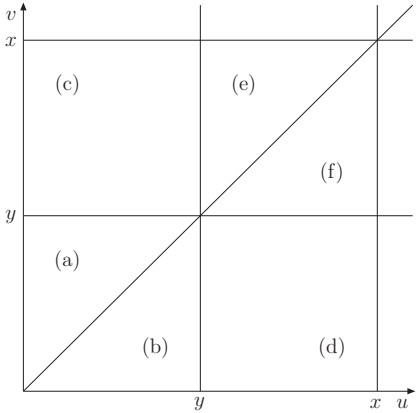}
\end{center}
\caption{\label{integration}The integration plane in the case $x^0>y^0$.}
\end{figure}
For the remaining terms integration over the contour $\cal C$ reduces 
to a ``single time'' integration over a combination of the six regions 
in Fig.\,\ref{integration}. Note that the upper limits of the integration
never exceed the largest time argument ($x^0$ in Fig.\,\ref{integration})
which ensures the causality of the Kadanoff--Baym equations. In most 
cases integration over a part of the $uv$ plane can be easily represented 
as integration over the whole plane, $0<u,v<\infty$, if two of the spectral 
functions are replaced by the corresponding retarded and (or) advanced 
propagators. There are, however, two exceptions: if the resulting integral is 
only over region $(e)$ or region $(f)$ in Fig.\,\ref{integration}. Using 
the identities  
\begin{subequations}
\begin{align}
\textint_{(e)} &=\textint_{(a+c+e)} +\textint_{(b)} -
\textint_{(a+b+c)} \,,\\
\textint_{(f)}  &=\textint_{(b+d+f)} +\textint_{(a)} -
\textint_{(a+b+d)} \,,
\end{align}
\end{subequations}
and the definitions of the retarded and advanced propagators, we can 
represent the corresponding contributions as combinations of integrals 
over the whole $uv$ plane. Collecting all the terms, we obtain expressions 
presented in Eqs.\,\eqref{Sigmad} and \eqref{Pid}.

\section{\label{WignerTrafo}Wigner transformation}
To calculate the self-energies entering the Boltzmann equations, 
we have to Wigner transform products of several two-point functions. Using 
the definitions \eqref{WignerTrafos}, we obtain for the Wigner transform 
of a product of $n$ functions of the same arguments \cite{Hohenegger:2008zk}:
\begin{align}
\label{xyWigner}
f_1(x,y&)\ldots  f_n(x,y) \rightarrow \textint d\Pi_{p_1} \ldots d\Pi_{p_n}(2\pi)^4\nonumber\\
&\times\delta^g (p-p_1-\ldots p_n){f}(X,p_1)\ldots {f}(X,p_n)\,.
\end{align}
Equation \eqref{xyWigner} allows us to Wigner transform the self-energies 
\eqref{Sigmab}, \eqref{Sigmac} and \eqref{Pic}. 
The self-energy  \eqref{Sigmad} has a more complicated structure:
\begin{align}
\label{Sigmastruct}
f(x,y)&=\textint_{vu} f_1(y,v)f_2(u,v)f_3(u,x)f_4(y,u)f_5(x,v)\,.
\end{align}
We will now calculate the Wigner transform of \eqref{Sigmastruct} in the 
Boltzmann approximation. That is, in each $f_n$ we will neglect the 
deviation of the corresponding center coordinate from $X\equiv X_{xy}$.
For instance:
\begin{align}
f_1(y,v)\rightarrow f_1(X_{yv},s_{yv}) \rightarrow  f_1(X_{xy},s_{yv})\,.
\end{align}
In this approximation the integration over $u$ and $v$ induces 
two conditions on the momenta: $p_v=p_u=0$, where $p_u=p_2+p_3-p_4$
and $p_v=p_1+p_2+p_5$. Integration over the relative coordinate $s$, 
see Eq.\,\eqref{DFXp}, induces an additional constraint: $p=p_s$, where 
$p_s={\textstyle\frac12}(p_5-p_4-p_3-p_1)$. Thus, in the Boltzmann 
approximation the Wigner-transform of \eqref{Sigmastruct} takes the
form:
\begin{align}
\label{SigmaWignTrafo}
{f}(X,p)&=\textint d\Pi_{p_1}\ldots\,d\Pi_{p_5}(2\pi)^{4}
\delta^g(p_u)(2\pi)^{4}\delta^g(p_v)\nonumber\\
&\times (2\pi)^{4}\delta^g(p-p_s) \,{f}_{1}(X,p_1)
\ldots\,{f}_{5}(X,p_5)
\end{align}
As far as decays are concerned, two of the momenta in \eqref{SigmaWignTrafo}
correspond to the initial and  final states, whereas three of the 
momenta correspond to the internal lines of the loop. The Dirac
$\delta$ functions in \eqref{SigmaWignTrafo} ensure conservation of four-momentum in each vertex
of the loop.  

The self-energy \eqref{Pid} has the structure
\begin{align}
f(x,y)&=\textint_{vu} f_1(v,u)f_2(x,v)f_3(x,u)f_4(y,v)f_5(y,u)\,.
\end{align}
Proceeding in the same way, we again obtain \eqref{SigmaWignTrafo} but now 
with $p_v=p_1-p_2-p_4$, $p_u=p_1+p_3+p_5$ and 
$p_s={\textstyle\frac12}(p_2+p_3-p_4-p_5)$. This completes the calculation
of the Wigner transforms of the self-energies.

\section{\label{Kinematics}Kinematics of the decay}
In this appendix, we discuss the kinematics of  the decay $\psi_i\rightarrow bb$ in the rest (in the early universe -- comoving) frame of the 
thermal bath and in the rest frame of the decaying heavy particle. For
simplicity, we assume that the masses of the toy baryons are negligibly
small in comparison with the mass of the heavy real scalar.
\begin{figure}[!ht]
	\begin{center}
	\includegraphics[width=0.48\textwidth]{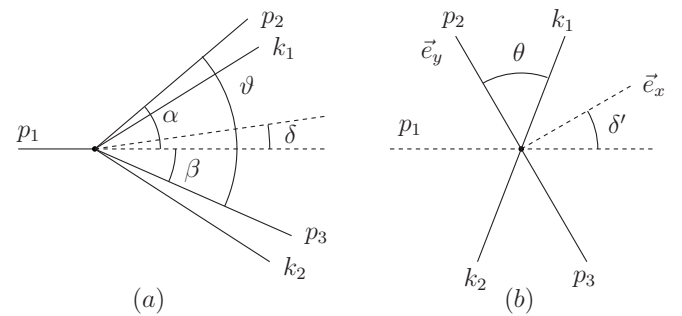}
	\end{center}
\caption{\label{restframedecay}Decay of the heavy scalar in the rest frame 
of the plasma and in the rest frame of the decaying particle (see
also Fig.\,\ref{cpcontributions} for the assignment of the different
momenta).}
\end{figure} 
Let us first consider  the  decay in the rest frame of the thermal bath. 
Energy-momentum conservation
tells us that the momentum $p_1$ of the  decaying particle
and the momenta $p_2$, $p_3$ of the  decay products must 
lie in the same plane. The latter depend on the 
scattering angle $\vartheta$, see Fig.\,\ref{restframedecay}a. 
Denoting the angle between the bisectrix
of the scattering angle and the momentum $\vec{p}_1$ of the decaying
particle by $\delta$ we obtain
\begin{align}
\label{thetadelta}
\vartheta=2\arccos\left(\frac{|\vec{p}_1|}{E_1}\cos\delta\right)\,.
\end{align}
The angles between the momentum of the decaying particle and those of 
the decay products are given by $\alpha=\vartheta/2+\delta$ and 
$\beta=\vartheta/2-\delta$. If $\delta=0$ then $\alpha=\beta$.
If $\delta\rightarrow \pm\frac{\pi}{2}$, then $\alpha\rightarrow \pi$ 
and $\beta\rightarrow 0$ and vice versa. Energy-momentum conservation 
implies, that the energies of the decay product are related to the angle 
$\delta$ by 
\begin{align}
\label{energies}
E_{2,3}=\frac{E_1}{2}\left(
1\mp\frac{|\vec{p}_1| \sin\delta}{\sqrt{E_1^2-|\vec{p}_1|^2\cos^2\delta}}
\right)\,.
\end{align}
If $\vec{p}_1=0$, then the energy is equally distributed between the 
decay products. This is also the case if $\vec{p}_1\neq 0$ but $\delta=0$.  
In any other case the energy is distributed unequally. In particular 
if $\delta\rightarrow \frac{\pi}{2}$, then $E_{2,3}=\frac12(E_1\mp |\vec{p}_1|)$,
so that in the ultrarelativistic limit one of the decay products has 
almost zero energy, whereas the other receives almost all energy of the 
decaying particle. 

As follows from Eq.\,\eqref{epsilon0}, to calculate the \textit{CP}-violating parameter, 
we need to evaluate the distribution functions associated with the statistical
propagators contributing to the vertex
loop. As discussed in Sec.\,\ref{CPviol}, only those  two terms of
\eqref{epsilon0} for which  both intermediate toy baryons are on-shell, whereas the real scalar
is off-shell, contribute to the \textit{CP}-violating
parameter (see Fig.\,\ref{cpcontributions}a). Since both momenta $k_1$ and $k_2$
of the internal toy baryon lines are on-shell, the kinematics of these
intermediate states is the same as the kinematics of the decay products. However, 
due to the presence of the intermediate off-shell real scalar, the corresponding
scattering angle does not  need  to be equal to $\vartheta$ and the angle between 
the two scattering planes can differ from zero. It is somewhat easier to perform 
the simultaneous analysis of the intermediate and final states' kinematics in the
rest frame of the decaying particle. The Lorentz transformation between the two 
frames reads
\begin{align}
\label{LorentzTrafo}
\hat{\Lambda}=\frac{1}{M}\left(
\begin{tabular}{cc}
 $E_1$ & $-|\vec{p}_1|$ \\
 $-|\vec{p}_1|$ & $E_1$
\end{tabular} 
\right).
\end{align}
As follows from Eq.\,\eqref{thetadelta}, in the new frame the scattering 
angle is $\vartheta'=\pi$ for both intermediate and final states, whereas 
Eq.\,\eqref{energies} implies that the energies are equal to $M/2$.
Using the fact  that components of the momentum orthogonal to the  direction 
of the boost 
are invariant under transformation \eqref{LorentzTrafo}, we can calculate 
the angle $\delta$ in the new frame: 
\begin{align}
\label{deltaprime}
\sin\delta'=\frac{E_1\sin\delta}{\sqrt{E_1^2-|\vec{p}_1|^2\cos^2\delta}}=
(|\vec{p}_3|-|\vec{p}_2|)/|\vec{p}_1|\,.
\end{align}
It then follows that $\delta'>\delta$ since the denominator of \eqref{deltaprime} 
is smaller than $E_1$.

For a homogeneous and isotropic system the one-particle distribution
functions depend only on the Lorentz-invariant product $ku$, where 
$k$ is the particles' momentum and $u$ is the four-velocity of the thermal
bath's rest frame with respect to the chosen frame of reference. In 
particular in thermal equilibrium: 
\begin{align}
f^{eq}(k)= [\exp((ku-\mu)/T)-1]^{-1} \,.
\end{align}
In the rest frame of the gas $u=(1,0,0,0)$, and we recover the usual 
Bose--Einstein distribution. Applying the Lorentz transformations 
\eqref{LorentzTrafo}, we can deduce $u$ in the rest frame of the decaying
particle 
\begin{align}
u=M^{-1} (E_1,-\vec{p}_1)\,.
\end{align}
Introducing an orthogonal coordinate system, as is depicted in Fig. 
\ref{restframedecay}, we can then express the arguments of the 
distribution functions in the form:
\begin{align}
\hspace*{-1.5mm}
uk_{1,2}={\textstyle\frac12}[E_1+|\vec{p}_1|(\sin\theta\cos\varphi\cos\delta' \mp 
\cos\theta\sin\delta')],
\end{align}
where $\varphi$ is the angle between the scattering planes of the intermediate
and final states (not depicted in Fig.\,\ref{restframedecay}).

\section{\label{app:ThermalAv}Thermal average of the \CP-violating parameter}
The \CP-violating parameter in vacuum is momentum-independent due to Lorentz invariance.
Since the surrounding medium defines a preferred frame (its rest frame), the effective
\CP-violating parameter in medium depends explicitly on the momenta of the participating
particles (see Fig.\,\ref{cpcontributions}). In order to investigate the order of
magnitude of the medium corrections, we consider the thermally averaged
\CP-violating parameter:
\begin{align}
	\label{epsthermaverage}
	\langle\epsilon_i\rangle = \frac{ \int d\Pi^3_{p_1} d\Pi^3_{p_2} d\Pi^3_{p_3} w(p_1,p_2,p_3) 
	\epsilon_i(p_1,p_2) }{ \int d\Pi^3_{p_1} d\Pi^3_{p_2} d\Pi^3_{p_3} w(p_1,p_2,p_3) }\,,
\end{align}
where  $\epsilon_i(p_1,p_2)=\epsilon_i^{\it vac}+\delta\epsilon_i^{\it med}(p_1,p_2)$ and $w$ represents the gain or loss term (they are equal in equilibrium). For 
the decay processes:
\begin{align}
	\label{w}
	w(p_1,p_2,p_3) & =  (2\pi)^4 \delta(E_1-E_2-E_3)\delta(\vec{p}_1-\vec{p}_2-\vec{p}_3)\nonumber \\
	& \times f(\vec p_2) \bar{f}(\vec p_3) [1+f_{\psi_i}(\vec p_1)]\,.
\end{align}
In the hierarchical limit $M_1 \ll M_2$, the \CP-violating parameter $\epsilon_1$ 
only depends on $|\vec{p}_1|$, see Eq.\,\eqref{epsilon hierarchical massless}. Furthermore, 
we set $m\rightarrow 0$ here. 
Integrating in \eqref{epsthermaverage} over momenta 
of the  final states, we find in this approximation:  
\[
	\langle\epsilon_1\rangle = \frac{ \int_0^\infty dq \; \omega(q) \epsilon_1(q) }{ \int_0^\infty dq \; \omega(q) } \;,
\]
where $q=|\vec{p}_1|$. To calculate the weighting 
function $\omega(q)$ we insert thermal distributions characterized by a 
common temperature $T$ and zero chemical potential.
We consider two cases: 
\begin{itemize}
	\item[(i)] in the first case  we insert Bose--Einstein (BE) distributions 
		for $f$ and $f_{\psi_i}$ in \eqref{w};
	\item[(ii)] in the second case we insert Maxwell--Boltzmann (MB) distribution
		for $f$ and neglect $f_{\psi_i}$ in \eqref{w}.
\end{itemize}
The resulting expressions for the weighting function read:
\begin{align}
	w(q) = \left\{
		\begin{array}{ll}
			\frac{2q}{E_q\sinh^2\left(\frac{E_q}{2T}\right)}\ln\left( \frac{ \sinh\left( \frac{ E_{q} + q }{ 4T } \right) }
			{  \sinh\left( \frac{ E_q - q }{ 4T } \right) } \right) & \quad \mbox{BE}\; , \\
			q^2 \exp\left(-\frac{E_q}{T}\right)
			& \quad \mbox{MB}\; , \\
		\end{array} \right.
\end{align}
where $E_q=\sqrt{M_1^2+q^2}$.

\section{Numerical details}\label{app:NumericalDetails}

To solve the system of Boltzmann equations (\ref{QuantumBoltzmannEquations}),
we introduce the transformed variables $x=a(t)$ and $\momitrans=Sa(t)\momiabs$
for time and momentum. The constant factor $S$ is chosen such that
$Sx=T^{-1}=(2\cdot 1.66 \sqrt{g_*}/M_{\text{Pl}}t)^\frac12$. The distributions
as functions of the transformed momenta are then well represented, in some sense,
in the range $\momitrans\simeq\Clowerbound-\Cupperbound$.\footnote{In particular
we require that the approximate numerical value of the moments
(\ref{number density and energy density}) are close to their true values for
close-to-equilibrium distributions. Also we demand that particles created
in decays are not produced with momenta outside of this range to a significant
extent so that total number densities show the expected behavior.} In addition, we introduce
the transformed on-shell energies and masses, $\massitrans =SxM_i$ and
$\enitrans =({\momitrans}^2 +{\massitrans}^2)^\frac12=Sx({\momiabs}^2 +{M_i}^2)^\frac12$.
In these coordinates the Liouville operator for Robertson--Walker space-time takes the form 
$\lorig [f](\momkabs)\rightarrow S^{-1}\hubblerate \enktrans \,\partial \ftrans{}{} (\momktrans)/\partial x$,
where $\ftrans{}{}(\momktrans)$ is the transformed one-particle distribution function
dependent on $\momktrans$ and $x$. Defining
$\ltrafo [\ftrans{}{}](\momktrans)\equiv \partial \ftrans{}{}(\momktrans)/\partial x$, the Boltzmann
equations can be written in the form $\ltrafo [\ftrans{}{}](\momktrans) = \carray{}{}{.\ftrans{}{}.}(\momktrans)$
with  transformed collision integral $\carray{}{}{.\ftrans{}{}.}$.

\begin{figure}[t]
	\begin{center}
		\includegraphics[width=\columnwidth]{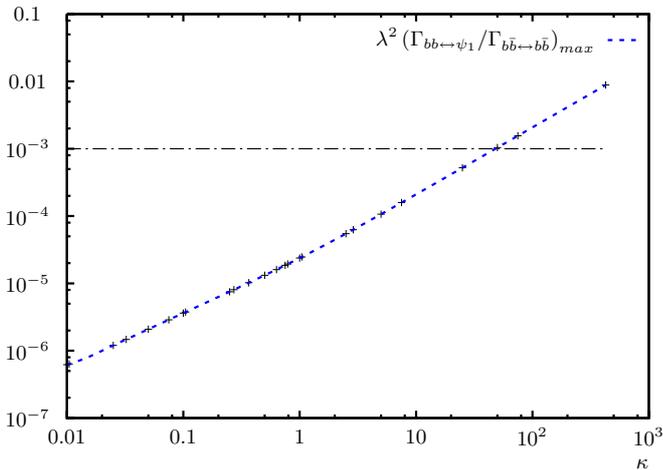}
	\end{center}
	\caption{\label{ratesratiomaxvswashout} The maximum value of the ratio of the rates 
	for $bb\leftrightarrow\psi$ and $b \bar{b}\leftrightarrow b \bar{b}$ over washout factor $\kappa$.}
\end{figure}

\begin{figure}[t]
	\begin{center}
		\includegraphics[width=\columnwidth]{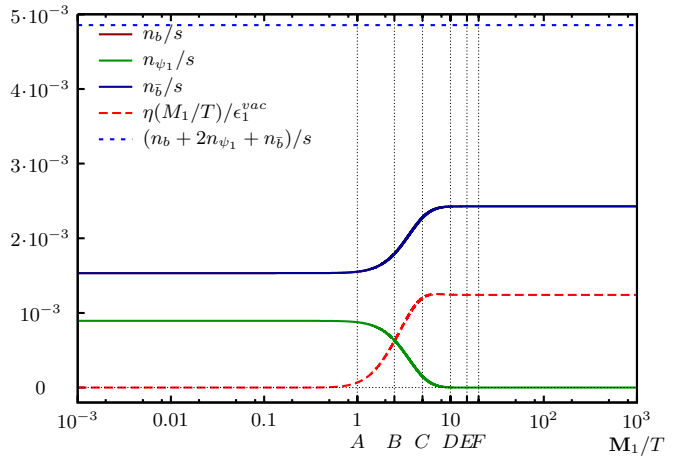}
	\end{center}
	\caption{\label{n_19_36_06} Number densities of the various species and the generated asymmetry 
	$\eta$ as functions of $M_1/T$ for $\kappa\simeq\Ckappac$ (case $c$). The total number density 
	is approximately conserved.}
\end{figure} 

\begin{figure}[t]
	\begin{center}
	\includegraphics[width=\columnwidth]{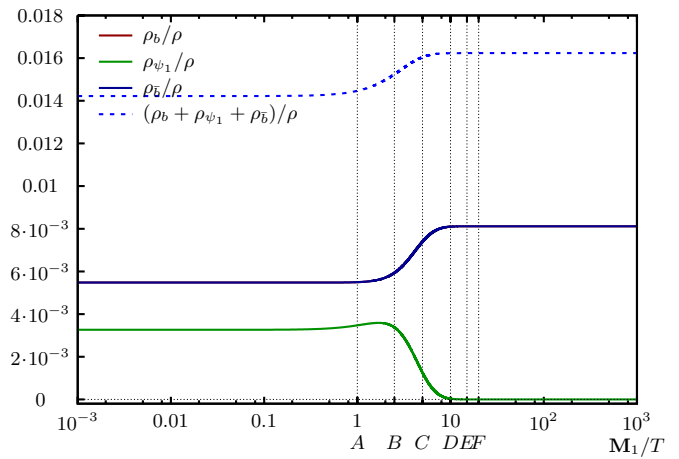}
	\end{center}
	\caption{\label{en_19_36_06} Energy densities of the various species as functions of $M_1/T$ 
	for $\kappa\simeq\Ckappac$ (case $c$). The ratio of the total energy density 
	$\rho_{b} + \rho_{\bar{b}} + \rho_{\psi}$ and the total cosmological energy density $\rho$ is 
	not constant. This feature is due to the different scaling behavior of relativistic and 
	nonrelativistic species. For this reason the ratio $\rho_\psi/\rho$ increases slightly before the 
	particles start to decay. This is more pronounced for smaller washout factors (see Fig.\,\ref{en_02_06_13}).}
\end{figure} 

\begin{figure}[t]
	\begin{center}
		\includegraphics[width=\columnwidth]{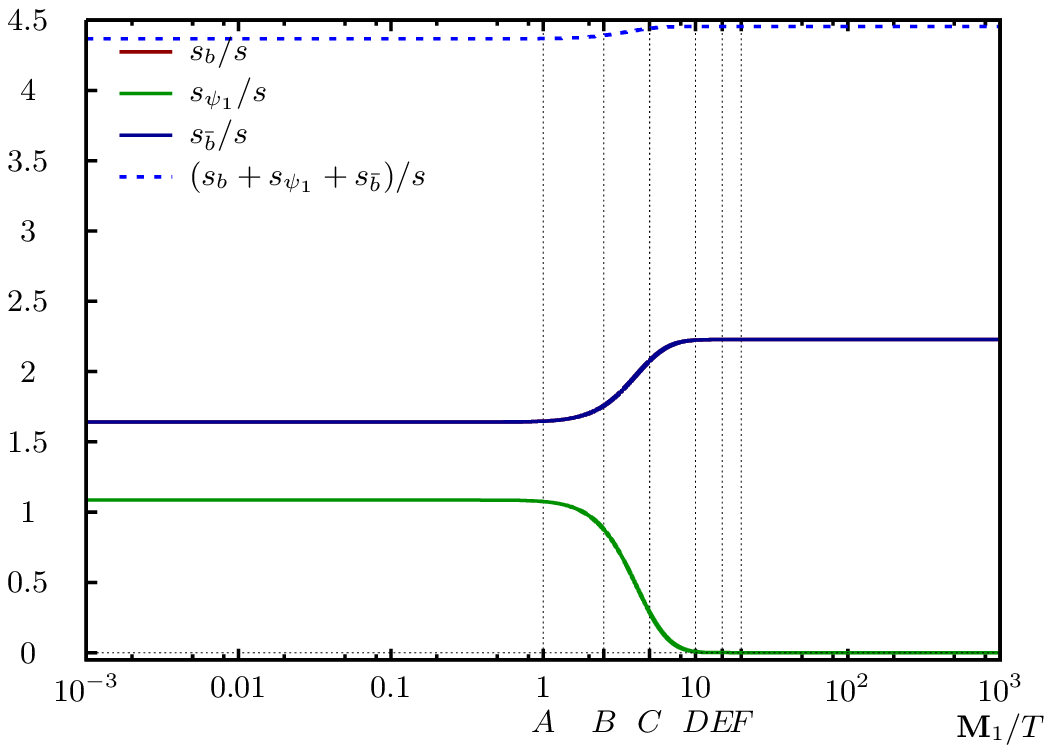}
	\end{center}
	\caption{\label{s_19_36_06} Entropy densities of the various species and the total entropy 
	density $(s_{b} + s_{\bar{b}} + s_{\psi})/s$ as functions of $M_1/T$ for $\kappa\simeq\Ckappac$ (case $c$).}
\end{figure} 

\begin{figure}[t]
	\begin{center}
		\includegraphics[width=\columnwidth]{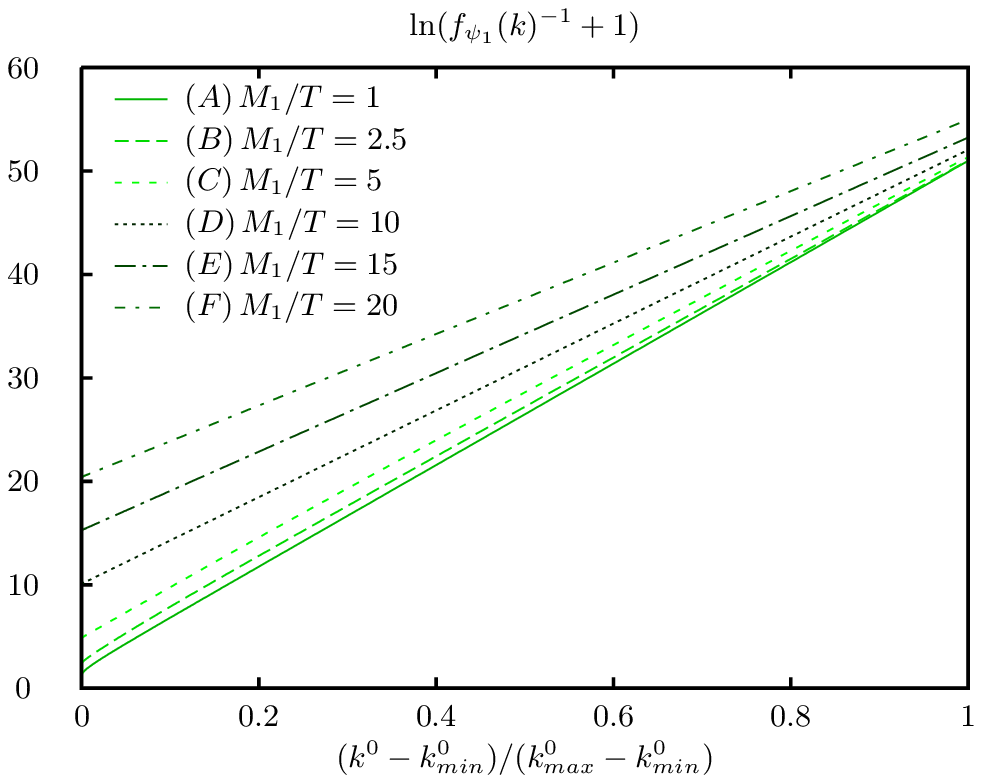}
	\end{center}
	\caption{\label{lnfvsen_19_36_06} Deviation of the distribution function $f_{\psi_1}$ from 
	equilibrium for washout factor $\kappa\simeq\Ckappac$ (case $c$).}
\end{figure} 

\begin{figure}[t]
	\begin{center}
	\includegraphics[width=\columnwidth]{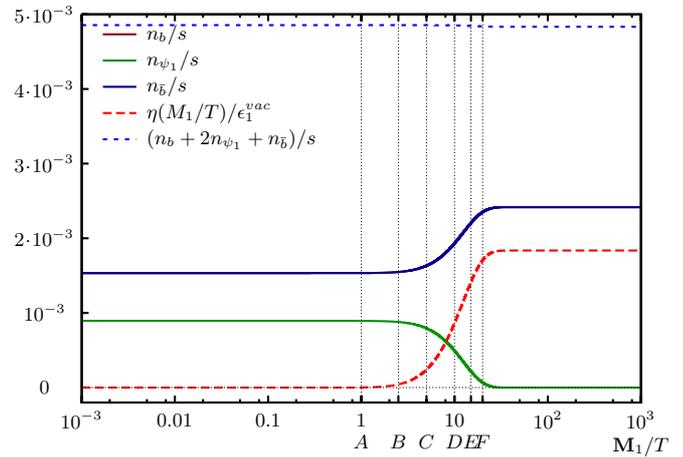}
	\end{center}
	\caption{\label{n_02_06_13} Number densities of the various species and the generated asymmetry 
	$\eta$ as functions of $M_1/T$ for $\kappa\simeq\Ckappaa$ (case $a$).}
\end{figure} 

\begin{figure}[t]
	\begin{center}
		\includegraphics[width=\columnwidth]{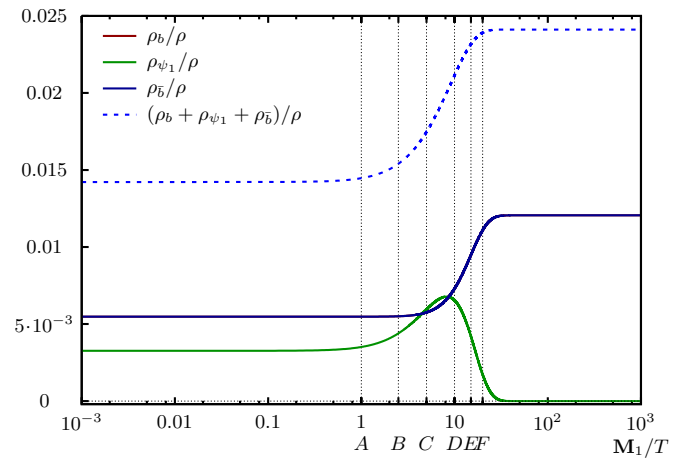}
	\end{center}
	\caption{\label{en_02_06_13} Energy densities of the various species and the total energy density 
	$(\rho_{b} + \rho_{\bar{b}} + \rho_{\psi})/\rho$ as functions of $M_1/T$ for $\kappa\simeq\Ckappaa$ (case $a$).}
\end{figure} 

\begin{figure}[t]
	\begin{center}
		\includegraphics[width=\columnwidth]{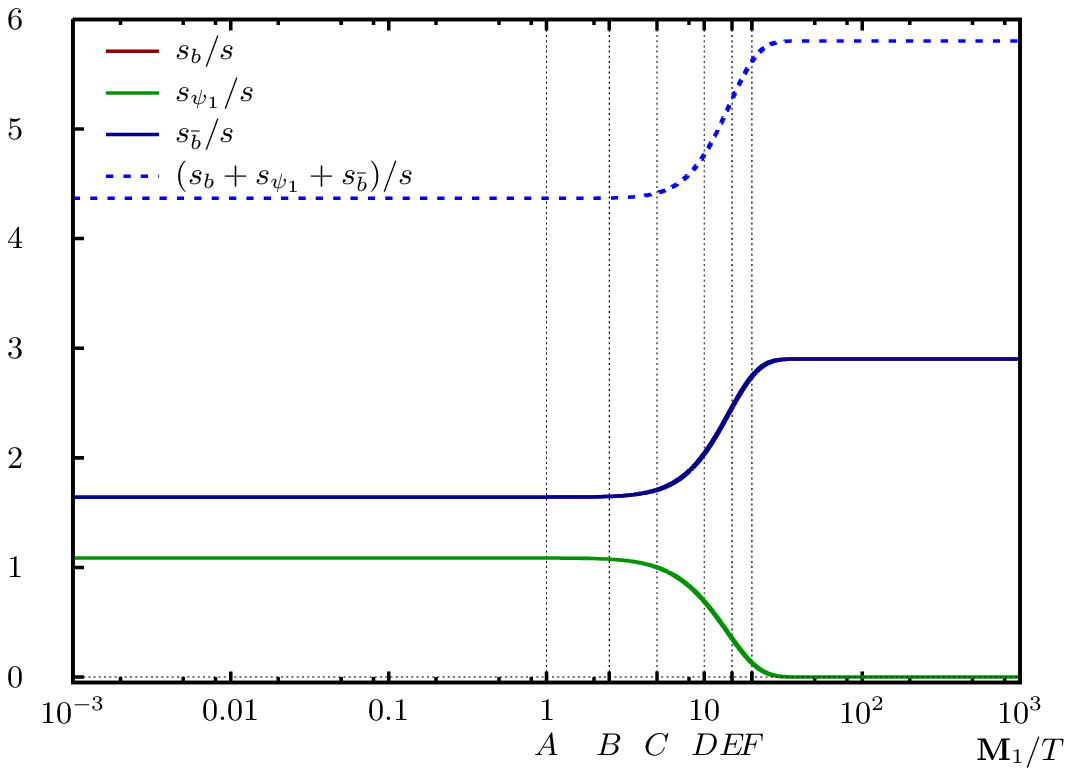}
	\end{center}
	\caption{\label{s_02_06_13} Entropy densities of the various species and the total entropy 
	density $(s_{b} + s_{\bar{b}} + s_{\psi})/s$ as functions of $M_1/T$ for $\kappa\simeq\Ckappaa$ (case $a$).}
\end{figure} 

\begin{figure}[t]
	\begin{center}
		\includegraphics[width=\columnwidth]{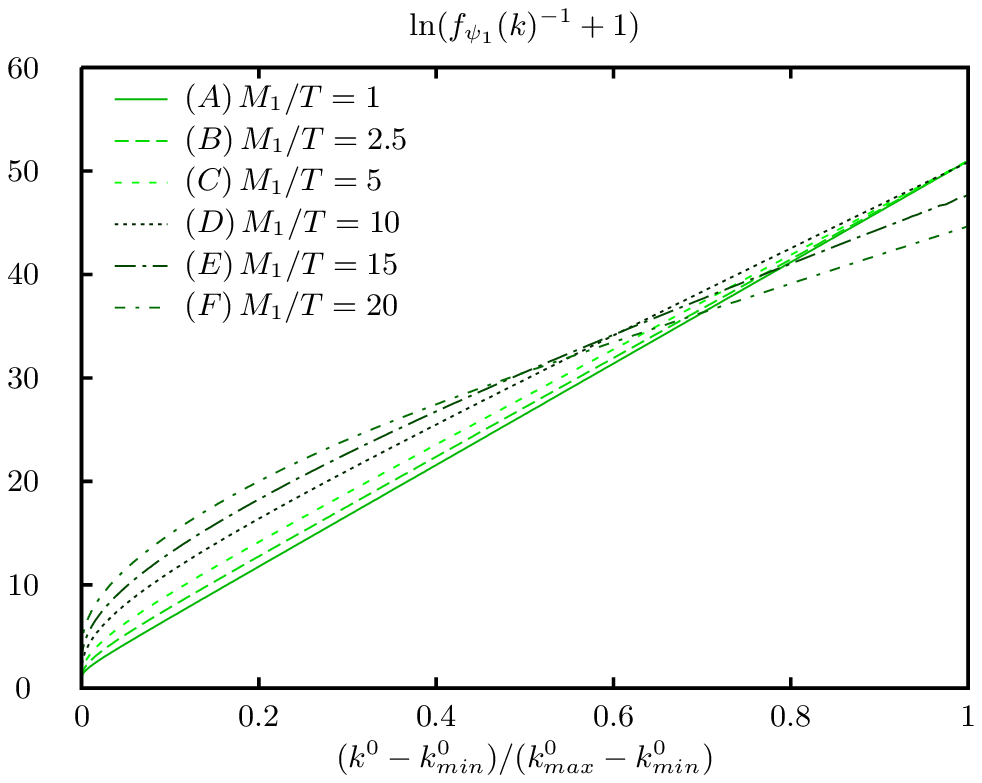}
	\end{center}
	\caption{\label{lnfvsen_02_06_13} Deviation of the distribution function $f_{\psi_1}$ from 
	equilibrium for washout factor $\kappa\simeq\Ckappaa$ (case $a$). \protect}
\end{figure}

Homogeneity and isotropy can be exploited to simplify the collision integrals significantly. 
In \cite{hohenegger:063502}, it has been shown how the various collision terms for decays, 
inverse decays and $2-2$ scattering can be reduced to lower dimensional integrals in general. 
Here, we transform the integrals to the new coordinates at the same time.
In particular, the collision integrals for a scattering process $12\leftrightarrow 34$ 
(here $b\bar{b}\leftrightarrow b\bar{b}$, $bb\leftrightarrow bb$ and $\bar{b}\bar{b}\leftrightarrow \bar{b}\bar{b}$) 
can be reduced to a twofold integral:
\begin{align}
	&\carray{12\leftrightarrow 34}{\momktrans}{.\ftrans{1}{}.}=\nonumber\\
	&{\textstyle \frac{1}{S\hubblerate x^2}\frac{1}{64\pi^3\enktrans}}\textint\textint 
		\frac{\momqtrans\dif \momqtrans}{\enqtrans}\frac{\momrtrans\dif \momrtrans}{\enrtrans} 
		\heaviside{\enptrans -\massptrans}D_{12\leftrightarrow34}\times\nonumber\\
	&\{[1+\ftrans{1}{}(\momktrans)][1+\ftrans{2}{}(\momptrans)]\ftrans{3}{}(\momqtrans)\ftrans{4}{}(\momrtrans)-\nonumber\\
	&\ftrans{1}{}(\momktrans)\ftrans{2}{}(\momptrans)[1+\ftrans{3}{}(\momqtrans)][1+\ftrans{4}{}(\momrtrans)]\}\,,
\end{align}
where $\enptrans = \enqtrans+\enrtrans-\enktrans$ and $\momptrans=[(\enptrans)^2 -{\massptrans}^2]^\frac12$.
The integrated scattering kernel $D_{12\leftrightarrow 34}$ for a constant 
(momentum-independent) 
amplitude $\mathcal{A}$ and for massless species $1$, $2$, $3$ and $4$  is given by
\begin{align}
	&D_{12\leftrightarrow 34}(\momktrans,\momptrans,\momqtrans,\momrtrans)=
		{\textstyle \frac{\mathcal{A}}{2\momktrans}}\heaviside{\momqtrans+\momrtrans - \abs{\momktrans-\momptrans}}\times\nonumber\\
	&\heaviside{\momktrans+\momptrans-\abs{\momqtrans-\momrtrans}}(\momqtrans+\momrtrans-
		\abs{\momqtrans-\momktrans}-\abs{\momrtrans-\momktrans})\,.
\end{align}
Similarly, the collision integrals for a particle created in inverse decays, $1\leftrightarrow 23$ 
(here $\psi_1\leftrightarrow bb$ and $\psi_1\leftrightarrow \bar{b}\bar{b}$), can be reduced to a single integral:
\begin{align}
	\carray{1\leftrightarrow 23}{\momktrans}{.\ftrans{1}{}.}=&{\textstyle \frac{S}{\hubblerate }
		\frac{1}{32\pi\enktrans}}\textint \frac{\momqtrans\dif \momqtrans}{\enqtrans}\heaviside{\enptrans-\massptrans}
		D_{1\leftrightarrow 23}\times\nonumber\\
	&\{[1+\ftrans{1}{}(\momktrans)]\ftrans{2}{}(\momptrans)\ftrans{3}{}(\momqtrans)-\nonumber\\
	&\ftrans{1}{}(\momktrans)[1+\ftrans{2}{}(\momptrans)][1+\ftrans{3}{}(\momqtrans)]\}\,,
\end{align}
where $\enptrans =\enktrans -\enqtrans$ and $\momptrans =[(\enptrans)^2 -{\massptrans}^2]^\frac12$. 
The integrated scattering kernel $D_{1\leftrightarrow 23}$ is given by
\begin{align}
	&D_{1\leftrightarrow 23}(\momktrans,\momptrans,\momqtrans)=\nonumber\\
	&=\frac{2\mathcal{A}}{\momktrans}\heaviside{\momktrans-\abs{\momptrans-\momqtrans}}
		\heaviside{(\momptrans+\momqtrans)-\momktrans}\,.
\end{align}
Finally, the collision integrals for a particle created in decays, $12\leftrightarrow 3$ 
(here $bb\leftrightarrow\psi_1$ and $\bar{b}\bar{b}\leftrightarrow\psi_1$), can be reduced to the single integral 
\begin{align}
	\carray{12\leftrightarrow 3}{\momktrans}{.\ftrans{1}{}.}=&{\textstyle \frac{S}{\hubblerate }\frac{1}{32\pi\enktrans}}
		\textint \frac{\momqtrans\dif \momqtrans}{\enqtrans}\heaviside{\enptrans-\massptrans}D_{12\leftrightarrow 3}\times\nonumber\\
	&\{[1+\ftrans{1}{}(\momktrans)][1+\ftrans{2}{}(\momptrans)]\ftrans{3}{}(\momqtrans)-\nonumber\\
	&\ftrans{1}{}(\momktrans)\ftrans{2}{}(\momptrans)[1+\ftrans{3}{}(\momqtrans)]\}\,,
\end{align}
where $\enptrans =\enqtrans -\enktrans$ and $\momptrans =[(\enptrans)^2 -{\massptrans}^2]^\frac12$. 
The integrated scattering kernel $D_{12\leftrightarrow 3}$ is given by
\begin{align}
	&D_{12\leftrightarrow 3}(\momktrans,\momptrans,\momqtrans)=\nonumber\\
	&=\frac{2\mathcal{A}}{{\momktrans}}\heaviside{{\momqtrans}-\abs{{\momktrans}-{\momptrans}}}
		\heaviside{({\momktrans}+{\momptrans})-{\momqtrans}}\,.
\end{align}
Number density and energy density corresponding to the distribution $\ftrans{}{}$ in transformed coordinates read
\begin{align}\label{number density and energy density}
&n[{\ftrans{}{}}]={\textstyle \frac{1}{2\pi^2}\left(\frac{1}{Sx}\right)^3}
\textint ({{\momktrans}})^2{\ftrans{}{}(\momktrans)}d  {{\momktrans}}\,,\nonumber\\
&\rho [{\ftrans{}{}}]={\textstyle \frac{1}{2\pi^2}\left(\frac{1}{Sx}\right)^4}
\textint ({{\momktrans}})^2{{{\enktrans}}}{\ftrans{}{}(\momktrans)}d  {{\momktrans}}\,.
\end{align}
For massless particles these are the second and third moment of the distribution, respectively.
As outlined in Sec.\,\ref{Numerics}, we assume that the interactions 
$bb\leftrightarrow bb$, $\bar{b}\bar{b}\leftrightarrow \bar{b}\bar{b}$ and 
$b\bar{b}\leftrightarrow b\bar{b}$ are rapid enough to keep the distribution functions 
of $b$ and $\bar{b}$ very close to their equilibrium distributions, parametrized by 
$a_0, a_1, \bar{a}_0$ and $\bar{a}_1$: 
\begin{align}\label{equilibrium distributions for b and bbar}
&f^{eq}_a(\momktrans)=[\exp(a_0+a_1\momktrans)-1]^{-1}\,,\nonumber\\
&\bar{f}^{eq}_a(\momktrans)=[\exp(\bar{a}_0+\bar{a}_1\momktrans)-1]^{-1}\,.
\end{align}
Assuming that $b\bar{b}\leftrightarrow b\bar{b}$ alone is much faster than the inverse decays 
into $\psi_1$, the evolution of $f$ and $\bar{f}$ can therefore be described by means of three 
parameters $a_0$, $\bar{a}_0$ and $a_1$. The equations for the evolution of these parameters 
are obtained by forming the moments $n[.]$ of Eqs.\,(\ref{boltzmann equation b}) and 
(\ref{boltzmann equation bbar}):\footnote{Here and in the following we use the abbreviations 
$\carray{bb\leftrightarrow \psi_1}{}{}=\carray{bb\leftrightarrow \psi_1}{}{\feqtrans{},\fpsitrans{}}$ and 
$\carray{\bar{b}\bar{b}\leftrightarrow \psi_1}{}{}=\carray{\bar{b}\bar{b}\leftrightarrow \psi_1}{}{\fbareqtrans{},\fpsitrans{}}$. 
Also note that $f$, $\bar{f}$ and $f_{\psi_1}$ are functions of the transformed coordinates, here.}
\begin{align}\label{parametrized equation one and two}
	n\left[\ltrafo [{\feqtrans{}}]\right]=&\frac{da_0}{dx}n\left[\frac{\partial{\feqtrans{}}}{\partial a_0}\right]+
		\frac{da_1}{dx}n\left[\frac{\partial{\feqtrans{}}}{\partial a_1}\right]=\nonumber\\
	&n\left[\carray{bb\leftrightarrow \psi_1}{}{}\right]\,,\nonumber\\
	n\left[\ltrafo [\fbareqtrans{}]\right]=&\frac{d\bar{a}_0}{dx}n\left[\frac{\partial\bar{\feqtrans{}}}{\partial \bar{a}_0}\right]+\frac{da_1}{dx}n\left[\frac{\partial\bar{\feqtrans{}}}{\partial a_1}\right]=\nonumber\\
	&n\left[\carray{\bar{b}\bar{b}\leftrightarrow \psi_1}{}{}\right]\,.
\end{align}
Here, we used $n\left[\carray{bb\leftrightarrow bb}{}{f}\right]=n\left[\carray{\bar{b}\bar{b}\leftrightarrow \bar{b}\bar{b}}{}{\bar{f}}\right]=0$ 
and $n\left[\carray{b\bar{b}\leftrightarrow b\bar{b}}{}{f,\bar{f}}\right]=n\left[\carray{\bar{b}b\leftrightarrow \bar{b}b}{}{\bar{f},f}\right]=0$.
The third equation is obtained by forming the moment $\rho[.]$ of the sum of Eqs.\, (\ref{boltzmann equation b}) and 
(\ref{boltzmann equation bbar}), i.e.
\begin{align}\label{parametrized equation three}
	&\rho\left[\ltrafo [\feqtrans{}]\right]+\rho\left[\ltrafo [\fbareqtrans{}]\right]=
		\frac{da_0}{dx}\rho\left[\frac{\partial \feqtrans{}}{\partial a_0}\right]+
		\frac{d\bar{a}_0}{dx}\rho\left[\frac{\partial \fbareqtrans{}}{\partial a_0}\right]+\nonumber\\
	&+\frac{da_1}{dx}\rho\left[\frac{\partial \feqtrans{}}{\partial a_1}\right]+\frac{d{a}_1}{dx}
		\rho\left[\frac{\partial \fbareqtrans{}}{\partial a_1}\right]=\nonumber\\
	&\rho\left[\carray{bb\leftrightarrow \psi_1}{}{}\right]+\rho\left[\carray{\bar{b}\bar{b}\leftrightarrow \psi_1}{}{}\right]\,,
\end{align}
where we used $\rho\left[\carray{b\bar{b}\leftrightarrow b\bar{b}}{}{f,\bar{f}}\right]+
\rho\left[\carray{\bar{b}b\leftrightarrow \bar{b}b}{}{\bar{f},f}\right]=0$.
The derivatives of $\feqtrans{}$ with respect to the parameters $a_i$ can be rewritten as
\begin{align}
	&\frac{\partial \feqtrans{{\momktrans}}}{\partial {a_i}} = 
		-({\enktrans})^i [1+\feqtrans{{\momktrans}}]\feqtrans{{\momktrans}}\,, i=0,1\,.
\end{align}
An analogous relation holds for the derivatives of $\fbareqtrans{}$ with respect to $\bar{a}_0$ 
and $a_1$. Solving Eqs. \eqref{parametrized equation one and two} and \eqref{parametrized equation three} 
for $da_0/dx$, $d\bar{a}_0/dx$ and $da_1/dx$, we find the differential equations for the three parameters:
\begin{align}
	&\frac{da_0}{dx}=-\frac{{\dot{a}_1}n\big[\enktrans (1+ \feqtrans{})\feqtrans{}\big]+
		n\big[\carray{bb\leftrightarrow \psi_1}{}{}\big]}{n\big[(1+ \feqtrans{})\feqtrans{}\big]}\,,\nonumber\\
	&\frac{d\bar{a}_0}{dx}=-\frac{{\dot{a}_1}n\big[\enktrans (1+ \fbareqtrans{})\fbareqtrans{}\big]+
		n\big[\carray{\bar{b}\bar{b}\leftrightarrow \psi_1}{}{}\big]}{n\big[(1+ \fbareqtrans{})\fbareqtrans{}\big]}\,,\nonumber\\
	&\frac{da_1}{dx} = -\Big(\big(n\big[\carray{bb\leftrightarrow \psi_1}{}{}\big]\rhobarf +
		n\big[\enktrans (1+\feqtrans{})\feqtrans{}\big]\rhobarc\big)\nonumber\\
 	&\times n\big[(1+\fbareqtrans{})\fbareqtrans{}\big] \rho\big[(1+\feqtrans{})\feqtrans{}\big]\nonumber\\
	&+\big(n\big[\carray{\bar{b}\bar{b}\leftrightarrow \psi_1}{}{}\big]\rhobarf+
		n\big[\enktrans (1+\fbareqtrans{})\fbareqtrans{}\big]\rhobarc\big)\nonumber\\
	&\times n\big[(1+\feqtrans{})\feqtrans{}\big] \rho\big[(1+\fbareqtrans{})
		\fbareqtrans{}\big]\Big)/\hab+\rhobarc/\rhobarf,
\end{align}
where we have defined
\begin{align}
	&\hab = \rhobarf \Big(n\big[(1+\feqtrans{})\feqtrans{}\big] n\big[(1+\fbareqtrans{})\fbareqtrans{}\big] \rhobarf \nonumber\\
	&+ n\big[\enktrans (1+\fbareqtrans{})\fbareqtrans{}\big] n\big[(1+\feqtrans{})\feqtrans{}\big] \rho\big[(1+\fbareqtrans{})\fbareqtrans{}\big]\nonumber\\
	&+ n\big[\enktrans (1+\feqtrans{})\feqtrans{}\big] n\big[(1+\fbareqtrans{})\fbareqtrans{}\big] \rho\big[(1+\feqtrans{})\feqtrans{}\big]\Big)\,,\nonumber\\
\end{align}
as well as 
\begin{align}
	\rhobarf =  & -\rho\big[\enktrans (1+\feqtrans{})\feqtrans{}\big]-\rho\big[\enktrans (1+\fbareqtrans{})\fbareqtrans{}\big]\,,\nonumber\\
	\rhobarc = & \rho\big[\carray{bb\leftrightarrow \psi_1}{}{}\big]+\rho\big[\carray{\bar{b}\bar{b}\leftrightarrow \psi_1}{}{}\big]\,.
\end{align}
As stated in the main text, we need to start with finite chemical potentials as to avoid the 
occurrence of Bose--Einstein condensation. We choose the minimal acceptable value $a_0=\bar{a}_0=0.5$, 
corresponding to $\mu_b =\mu_{\bar{b}} = -\Cmubinitial\,T_0$ and $\mu_{\psi_1}=2\mu_b$. 
The initial value $a_1=1$ corresponds to the initial cosmological temperature $T_0$.
We checked that the results do not depend on $T_0$ as long as $T_0 \gg M_1$.
The heavy species $\psi_1$ is subject to relatively weak interactions only, so that its distribution 
function can deviate from kinetic equilibrium. Therefore, we solve the full Boltzmann equation for $\psi_1$,
\begin{align}
	 \ltrafo [f_{\psi_1}](\momktrans) = \carray{\psi_1\leftrightarrow bb}{\momktrans}{f_{\psi_1},f}
		+\carray{\psi_1\leftrightarrow \bar{b}\bar{b}}{\momktrans}{f_{\psi_1},\bar{f}}\,,
\end{align}
along with the integrated ones for $b$ and $\bar{b}$.

Because of the integration of the equations for the massless species all collision terms for $2-2$ 
scattering drop out of the system. In order to verify that the rates for these processes are much 
larger than the ones of the decays and inverse decays we have computed the rates for these 
processes numerically. The maximum (during the full evolution) of the ratio of $\Gamma_{bb\leftrightarrow \psi_1}$ 
and  $\Gamma_{b\bar{b}\leftrightarrow b\bar{b}}$ is exemplarily presented in Fig.\,\ref{ratesratiomaxvswashout} 
(the rates for the other $2-2$ processes are similar). It shows that we can choose $\lambda\sim 1$ or smaller 
for most of the relevant range of $\left|g\right|^2$ if we demand that 
$\Gamma_{b\bar{b}\leftrightarrow b\bar{b}}/\Gamma_{bb\leftrightarrow \psi_1}\gtrsim \Cminrateratio$ 
as criterion that $b$ and $\bar{b}$ are in kinetic equilibrium at all times. Here the equilibrium shape of 
$f$ and $\bar{f}$ is not distorted by the expansion since we are dealing with massless particles. 
In addition, it can be argued that the $2-2$ processes are meant to model rapid gauge interactions with 
different particles which would have the same effect of equilibrating $b$ and $\bar{b}$. In this sense we 
could even formally tolerate nonperturbative values of $\lambda$.

To turn the equations into a system of ordinary differential equations (ODE) the distribution functions were discretized on a grid of 
dimension $\Cdim$ with linearly increasing spacings in the range $\momktrans\simeq\Clowerbound \ldots\Cupperbound$ 
to account for the characteristic behavior of close-to-equilibrium distributions at small and large 
momenta. All integrals were approximated by Riemann sums on this grid. The system of Boltzmann 
equations behaves numerically stiff. This means that it is advisable to use an implicit method for 
its numerical solution to achieve acceptable step sizes (and hence acceptable execution times and 
numerical errors). Here CVODE with its backward differentiation formula with Newton iteration was used 
as ODE solver. The full  Jacobian was computed analytically in every external step. A relative 
tolerance of $\Creltol$ was attributed to every momentum mode. Because of the implicit method all solutions 
were computed in $\mathcal{O}(10^3)$ steps.

Since the global systematic error due to the discretization cannot be computed within the method
the proper behavior of the system was tested by successive refinement of the grid and comparison 
of some of the macroscopic quantities with the theory predictions. For this purpose, we present 
two examples of the number densities $n_x$, the energy densities $\rho_x$ and the entropy densities
for the washout factors $\kappa\simeq\Ckappac$ (case $c$ in  Fig.\,\ref{n_19_36_06}-\ref{s_19_36_06}) and
$\kappa\simeq\Ckappaa$ (case $a$ in Fig.\,\ref{n_02_06_13}-\ref{s_02_06_13}). The total number density
$(n_b + 2n_{\psi_1} +n_{\bar{b}})/s$ is almost conserved (as discussed in Sec.\,\ref{Numerics}).
The ratio $(\rho_b +\rho_{\psi_1}+\rho_{\bar{b}})/\rho$ is not constant  (see Fig.\,\ref{en_19_36_06}). 
This behavior is expected for a system involving nonrelativistic massive particles and is also observed 
for the bottom-up equations. The ratio is much smaller than one so that it is justified to neglect the 
backreaction on the curvature. Finally, the total entropy density is steadily increasing as it should. 
Figures \ref{lnfvsen_19_36_06} and \ref{lnfvsen_02_06_13} show the deviation of the distribution 
function $f_{\psi_1}$ from kinetic equilibrium ones for which the curves would be straight lines. 
The deviation from equilibrium is larger for smaller values of $\kappa$ and increases at late times, as expected.

\end{appendix}



\end{document}